\documentclass[sigconf]{acmart}

\usepackage{multirow}
\usepackage{subcaption}
\usepackage{float}
\usepackage{soul,color}
\usepackage{colortbl}
\usepackage{amsmath}

\newcommand{\proposed}{CatAlyst}
\newcommand{\name}{Satoru}

\AtBeginDocument{%
  \providecommand\BibTeX{{%
    \normalfont B\kern-0.5em{\scshape i\kern-0.25em b}\kern-0.8em\TeX}}}

\copyrightyear{2023}
\acmYear{2023}
\setcopyright{rightsretained}
\acmConference[CHI '23]{Proceedings of the 2023 CHI Conference on Human Factors in Computing Systems}{April 23--28, 2023}{Hamburg, Germany}
\acmBooktitle{Proceedings of the 2023 CHI Conference on Human Factors in Computing Systems (CHI '23), April 23--28, 2023, Hamburg, Germany}\acmDOI{10.1145/3544548.3581133}
\acmISBN{978-1-4503-9421-5/23/04}

\newcommand{\tabref}[1]{Table~\ref{#1}}
\newcommand{\figref}[1]{Figure~\ref{#1}}
\newcommand{\secref}[1]{Section~\ref{#1}}

\newcommand{\etal}{\textit{et al.}}
\newcommand{\eg}{\textit{e.g.},~}
\newcommand{\ie}{\textit{i.e.},~}

\newcolumntype{C}[1]{>{\centering\arraybackslash}m{#1}}

\begin{document}

\title[\proposed: Domain-Extensible Intervention for Preventing Task Procrastination Using Large Generative Models]{\proposed: Domain-Extensible Intervention for Preventing Task Procrastination Using Large Generative Models}








\author{Riku Arakawa}
\orcid{0000-0001-7868-4754}
\affiliation{%
  \institution{Carnegie Mellon University}
  \city{Pittsburgh}
  \country{USA}
}
\email{rarakawa@cs.cmu.edu}
\authornote{These authors contributed equally and are ordered alphabetically.}

\author{Hiromu Yakura}
\orcid{0000-0002-2558-735X}
\affiliation{
    \institution{University of Tsukuba / National Institute of Advanced Industrial Science and Technology (AIST)}
    \city{Tsukuba}
    \country{Japan}
}
\email{hiromu.yakura@aist.go.jp}
\authornotemark[1]

\author{Masataka Goto}
\orcid{0000-0003-1167-0977}
\affiliation{
    \institution{National Institute of Advanced Industrial Science and Technology (AIST)}
    \city{Tsukuba}
    \country{Japan}
}
\email{m.goto@aist.go.jp}

\renewcommand{\shortauthors}{Arakawa and Yakura, et al.}



\begin{abstract} 

\proposed{} uses generative models to help workers' progress by influencing their task engagement instead of directly contributing to their task outputs.
It prompts distracted workers to resume their tasks by generating a continuation of their work and presenting it as an intervention that is more context-aware than conventional (predetermined) feedback.
The prompt can function by drawing their interest and lowering the hurdle for resumption even when the generated continuation is insufficient to substitute their work, while recent human-AI collaboration research aiming at work substitution depends on a stable high accuracy.
This frees \proposed{} from domain-specific model-tuning and makes it applicable to various tasks.
Our studies involving writing and slide-editing tasks demonstrated \proposed{}'s effectiveness in helping workers swiftly resume tasks with a lowered cognitive load.
The results suggest a new form of human-AI collaboration where large generative models publicly available but imperfect for each individual domain can contribute to workers' digital well-being.

\end{abstract}

\ccsdesc[500]{Human-centered computing~Interaction design}
\ccsdesc[500]{Computing methodologies~Machine learning}
\ccsdesc[300]{Human-centered computing~Natural language interfaces}

\keywords{large generative models, behavior change, task engagement, procrastination}

\maketitle

\section{Introduction}
\label{sec:intro}

Many office workers perform intellectual tasks, such as writing, coding, and editing slides, using computers every day.
Supporting workers through the development of intelligent systems is an important topic that has long been studied in the field of human-computer interaction (HCI), and a number of tools have been proposed, such as efficient editors, for each task.
An emerging trend in this context is to involve AI as a collaborative system.
The recent rapid development of generative models has enabled replacing some portion of human tasks.
For example, PEER~\cite{DBLP:journals/corr/abs-2208-11663} is a collaborative language model that can write drafts, add suggestions, and propose edits in writing.
GitHub Copilot~\cite{DBLP:conf/pldi/0001KLRRSSA22} is a pair programmer that can generate code corresponding to natural language descriptions of the target functionality.

While these systems that contribute to workers' tasks directly seem promising, their efficiency and subsequent workers' satisfaction depend on the accuracy and fidelity of the AIs.
For instance, trust may not be built between systems and workers because of poor outputs that conflict with workers' intentions or do not meet their expectations~\cite{Dietvorst_2015, DBLP:conf/ecis/JussupowBH20, DBLP:conf/chi/YinVW19, DBLP:journals/corr/abs-2204-08471}, resulting in ignorance of the systems.
For example, Yang \etal~\cite{DBLP:conf/iui/YangZZLL22} proposed an AI-based system that supports the writing of fictional stories through text generation; however, it was suggested that the output of the AI might irritate users due to its low accuracy, lack of usability, and insufficient transparency.
It is theoretically possible to address this accuracy issue by developing more accurate models (\eg by fine-tuning large generative models in each target domain).
However, it is not realistic to customize the models for each worker's task because of the huge cost of model training~\cite{DBLP:conf/aaai/StrubellGM20, DBLP:journals/corr/abs-2007-03051} in a variety of tasks and domains.

In contrast to tuning systems specific to each task (\eg writing and slide-editing), we propose a new approach, \textit{\proposed{}}, by designing AI systems that support workers' task engagement by inducing behavior changes using large generative models without tuning.
During an intellectual task, workers often experience a lack of progress and stay away from it because they lose focus owing to the high cognitive load required to perform the task~\cite{doi:10.1080/17470214808416738, 10.2307/1748692, Wang2020Achieving}.
Such behavior often leads to procrastination, causing stress and degraded self-efficacy~\cite{Lay1986Clarry, Frakes2016Procrastination, Saddler1993Multidimensional, Beheshtifar2011Effect}.
We introduce a design employing generative models to influence workers' behaviors when faced with such moments by lowering the cognitive load necessary to resume working on their tasks, as \textit{catalysts} in chemistry.
Specifically, \proposed{} first detects when workers' progress is halted and then intervenes with them by presenting the continuation of their work, which is generated by common large generative models.
Importantly, \proposed{} aims to retrieve their interest and encourage them to resume the original task by lowering the hurdle, not to contribute directly to workers' tasks with generated content.
We anticipate that current generative models can perform accurately enough to produce content to induce such behavior changes without any tuning for an individual task.
Moreover, since workers recognize \proposed{} as an intervention to encourage their work rather than proactively using AIs with high expectations, they would not feel as irritated as systems like~\cite{DBLP:conf/iui/YangZZLL22} when generative models perform poorly.
The following is a scenario describing how \proposed{} is expected to work.
\begin{quote}
    \name{} is a university student who works on a presentation assignment. He is unmotivated and has put off working on it until the deadline of three days. He finished roughly 70\% of the slide creation, but then no ideas came to mind, and he lost his concentration. He started watching popular drama on another tab of the browser. \proposed{} then detects his unengaged behavior and sends a notification on the browser, saying \textit{``I generated the continuation of the slide. Let's take a look at it''}. He paid attention to the notification while watching the drama, became interested in the generated content, and glanced at the slide-editing interface. The content was thought-provoking, although the formatting and flow were a little off from what \name{} had put. \name{} began to ponder further the direction of the updated slide. He then stopped watching the drama and resumed the slide-editing task by formatting the automatically created slide and reorganizing the entire structure. 
\end{quote}

In terms of increasing workers' engagement via intervention, several approaches have been proposed in the HCI literature.
For example, presenting an encouraging message or feedback is one useful approach in persuasive technology~\cite{DBLP:conf/lak/BernuyHSZLR0W22, DBLP:conf/ph/VriesZBDTE17, DBLP:conf/cscw/LiuJPP14}.
Previous studies have suggested that context awareness is important for effectiveness~\cite{DBLP:conf/hicss/RodriguezPB19}, and variation in the intervention content is no less essential for its continued effectiveness~\cite{DBLP:journals/pacmhci/KovacsWB18}.
\proposed{} follows these studies, as it aims at presenting more context-aware and variational interventions as a form of continuation of interrupted work.

To test the effectiveness of \proposed{}, we conducted a series of evaluations.
First, we developed a prototype running on a web page that supports writing tasks.
It generates a continuation of sentences in progress using GPT-3~\cite{DBLP:conf/nips/BrownMRSKDNSSAA20}, a large language model.
Our user study with 12 participants showed that \proposed{}'s intervention has two significant effects compared with a conventional intervention using encouraging messages: (1) reducing the time the participants spent recovering their interest in the task and (2) increasing their productivity after the intervention.
Furthermore, participants felt significantly less frustration while performing the task using \proposed{} and favored its usability.

Second, to examine how \proposed{} can be used in an unconstrained situation for a longer period, we conducted a user study in which ten participants used the prototype for five days.
The participants used the prototype for different purposes, such as keeping a diary, writing a novel as their hobby, and writing media articles as their job.
Overall, their comments after the trial reaffirmed the usefulness of \proposed{}, demonstrating different cases in which our system helped them with writing tasks in their lives.
Moreover, we found that they perceived \proposed{}'s benefits from various perspectives, shedding light on the role of \proposed{} as a reminder, an ideator, and a peer.

Finally, to demonstrate the extensibility of \proposed{} in different tasks, we developed a second prototype in the form of a Chrome extension that supports slide-editing tasks.
It generates the continuation of slides in progress using GPT-3~\cite{DBLP:conf/nips/BrownMRSKDNSSAA20} by providing slides in progress as a few-shot prompt.
It also incorporates a diffusion model~\cite{Rombach_2022_CVPR} to generate an image that complements the generated slide.
Our user study with 12 participants showed that \proposed{}'s intervention had two significant effects compared to a conventional intervention: (1) reducing the time the participants spent recovering their interest in the task and (2) reducing the time they spent on the task by increasing their productivity.
We also verified that \proposed{} posed less cognitive load to the participants, indicating the extensibility of our design.

While intelligent systems involving generative models tend to pursue their accuracy for each specific task to delegate larger parts of the tasks to AIs from humans, this is not always feasible owing to the issues with training data and computational costs.
This study shows that, under the condition that large generative models have a certain degree of generality but not optimal accuracy, it is possible to support workers in multiple tasks by lowering the hurdle for resumption.
Our results and discussion suggest a new form of human-AI collaboration that contributes to the digital well-being of office workers today.

\section{Related Work}
\label{sec:rw}

This study is placed in the context of human-AI collaboration, as it aims to support knowledge workers with AI techniques.
We first draw from the recent research trend of leveraging generative models to improve worker productivity while performing tasks.
Thereafter, we review and emphasize the need to support workers' task engagement to highlight the difference between our approach and previous studies.
Finally, we discuss factors affecting workers' task engagement based on existing HCI studies and introduce our idea of utilizing generative models for this purpose.

\subsection{Generative Models for Improving Task Efficiency}
\label{sec:rw-generative}

As many tasks have been digitized, there is a strong need for computational support to improve the efficiency of such tasks, producing many software systems designed for various tasks, such as Microsoft Office 365, Adobe Creative Cloud, \textit{etc}.
Such systems have now started to incorporate AI techniques to add machine intelligence to accelerate workers' task performance.
For example, Gmail~\cite{DBLP:conf/kdd/ChenLBCZLTWDCSW19} can automatically generate response candidates as well as complete sentences.
Adobe Photoshop~\cite{Adobe_Photoshop_2021} features neural filters that allow users to easily manipulate, for instance, people's hair color, emotion, and light condition.
GitHub Co-Pilot~\cite{DBLP:conf/pldi/0001KLRRSSA22} can generate programming code based on the natural language description of the functionality provided by users.
These products demonstrate the enormous potential of leveraging AI techniques to directly contribute to and replace parts of human tasks, significantly improving workers' task efficiency.

More features and interactions have been proposed at the research level using machine-learning-based generation techniques.
One of the most prominent examples is writing support systems that leverage large language models~\cite{DBLP:conf/nips/BrownMRSKDNSSAA20}.
There are a number of recent studies that demonstrate how human writers and AI-based generation techniques can collaborate to support writers~\cite{DBLP:conf/chi/0002LY22, DBLP:journals/corr/abs-2208-11663, DBLP:journals/corr/abs-2208-09323}, such as in creative writing~\cite{DBLP:conf/icids/RoemmeleG15, DBLP:conf/cscw/ShakeriND21, DBLP:conf/iui/YangZZLL22, Singh2022Where, DBLP:conf/iui/YuanCRI22}.
For example, Dang \etal~\cite{DBLP:journals/corr/abs-2208-09323} proposed a writing support tool that provides paragraph-wise summaries as self-annotations while workers write the content.
They found that the generated summaries gave the workers an external perspective on their writing and helped to polish the content.

Emerging generation techniques have also been proposed for tasks other than writing, such as slide-editing~\cite{DBLP:conf/kcap/Sefid0MG19, DBLP:conf/aaai/LiHML21, DBLP:conf/chi/ZhengWWM22}, music creation~\cite{DBLP:conf/chi/LouieCHTC20, DBLP:conf/chi/SuhYTC21}, and drawing~\cite{DBLP:conf/chi/LinGCYY20, DBLP:conf/iui/KimMS21, DBLP:conf/acmidc/ZhangZWHSLHYY21, DBLP:conf/chi/ZhangYWLLYY22}.
For example, Sefid \etal~\cite{DBLP:conf/kcap/Sefid0MG19} proposed a slide generation technique from PDF files for scientific papers, whereas Zheng \etal~\cite{DBLP:conf/chi/ZhengWWM22} proposed an approach of converting computational notebooks into slides and showed that it improved the efficiency of a slide-editing task.

While these studies show promising results of employing AI-based generation techniques to improve task efficiency, there are two gaps: the wall of accuracy and the dependency on worker engagement.
First, insufficient accuracy and fidelity in AI's output~\cite{DBLP:journals/corr/abs-2206-04615} can devastate trust in human-AI collaboration, leading to ignorance of such systems.
Yang \etal~\cite{DBLP:conf/iui/YangZZLL22} found that poor quality in generated texts could irritate fiction writers.
Kim \etal~\cite{DBLP:conf/iui/KimMS21} found that poor quality of output from AI does not provide rich inspiration for workers.
It would be ideal to customize models for each worker so that they can achieve sufficient accuracy in individual tasks~\cite{Sowa2021Cobots}; however, this is not necessarily feasible because of the huge cost of training and maintaining large models~\cite{DBLP:conf/aaai/StrubellGM20, DBLP:journals/corr/abs-2007-03051}.
Furthermore, Yang \etal~\cite{DBLP:conf/chi/YangSRZ20} pointed out the existence of inherent uncertainty surrounding their capabilities and output complexity, making it difficult to construct trustful systems, even with such customized models.
Given these inherent concerns, it is no less important to study generalizable approaches as to how generative models can contribute to workers performing various tasks without individual model tuning.

Second, the above studies implicitly assumed situations where workers are amenable to using the systems and continue working on tasks, which might not always hold in actual cases.
In fact, as we describe in the next section, workers frequently struggle to maintain focus while performing tasks for various reasons, including the required high cognitive load and multitasking.
When not only their attention is away from the task but also their motivation is low, workers might not fully benefit from the systems even if they feature more useful functions.

To overcome these two gaps, we propose a novel design of employing AI-based generation techniques to improve workers' task engagement without having to tune models to an individual task.
Our approach targets workers who may not be fully immersed in their task, which is an alternative to previous studies that aim to improve the productivity of workers who already engage in the task.
We discuss related work on computational support for workers' task engagement in the following subsections.

\subsection{Needs for Supporting Worker's Task Engagement}
\label{sec:rw-needs}

Modern knowledge workers need to concentrate on a task for a long time, such as writing, coding, and slide-editing.
However, it is difficult to maintain concentration for a long time because these tasks cause mental fatigue; for example, human concentration is known to break after 30--45 min when performing these tasks~\cite{doi:10.1080/17470214808416738, 10.2307/1748692}.
It is natural for people to become distracted, and this tendency is even higher in remote work scenarios~\cite{Wang2020Achieving}, which have increased significantly in recent years owing to COVID-19~\cite{Peixoto2021Role}.
Furthermore, since such tasks require a high cognitive load, workers often experience difficulty, particularly when starting tasks or resuming after interruption.
For example, Mark \etal~\cite{DBLP:conf/chi/MarkGK08} revealed that it could take 25 min to return to a task on average once distracted from the task.
This further leads to procrastination~\cite{Lay1986Clarry, Frakes2016Procrastination}, which is known to be a major problem for office workers today, as it can lead to stress and problems in relationships, work, and health, such as depression~\cite{Saddler1993Multidimensional, Beheshtifar2011Effect, DBLP:journals/pacmhci/ChoCKKCL21}.
Therefore, there is a crucial need to support workers dealing with these issues and increase their motivation and engagement in intellectual tasks.

Regarding workers' task engagement, it is known that the causes of the high hurdle to starting work are complex~\cite{DBLP:journals/corr/abs-2101-10191}.
One of the major causes is perfectionism among workers, leading to their assumption that tasks must be completed flawlessly.
They often stop thinking before starting tasks because they consider too much about the expected workload and sometimes feel fear of failure~\cite{Schouwenburg1992Procrastinators}.
In response, practitioners often emphasize the importance of starting tasks anyway, for instance, for 3--5 min before deeply thinking about the workload~\cite{Pychyl2013Solving}, as one of the best ways to engage in the tasks.
The Pomodoro technique~\cite{Cirillo2018Pomodoro} is a strategy to divide the entire task into small segments so that workers can start working on one of them with a lowered hurdle.

Our strategy is to design intervention by leveraging generative models to induce such behavior change, that is, making it easier for workers to take the first step to start working when they are not immersed in the task.
In the next subsection, we review existing work in the same context as supporting workers' task engagement and discuss how generative models could be a powerful approach to achieving it.

\subsection{Interventions for Supporting Worker's Task Engagement}
\label{sec:rw-engagement}

There are several approaches proposed to support workers' task engagement through interventions when they are distracted.
Site blockers are a typical example of explicit intervention since they force users to return to their intended task by blocking access to specific entertainment websites.
However, this is a primitive and static approach that increases the psychological burden on the user owing to its coercive power~\cite{DBLP:journals/pacmhci/KovacsWB18}, resulting in workers uninstalling the systems.

By contrast, several studies in HCI also design interventions that affect people's behavior less explicitly.
Arakawa and Yakura~\cite{DBLP:conf/chi/ArakawaY21} proposed \textit{mindless intervention} to draw students' attention to online classes based on the design of Mindless Computing~\cite{DBLP:conf/huc/AdamsCJC15}.
Since such mindless interventions are designed for situations that do not demand users' conscious awareness, they could be effective in our case to draw workers' attention to the task (\eg writing).
However, this assumption is not always applicable when we try to get users to actually begin the task, given the higher hurdle imposed, as discussed in relation to procrastination.

Nudge has been extensively studied~\cite{DBLP:journals/corr/abs-2011-12468, DBLP:conf/cscw/LiuJPP14, DBLP:conf/ph/VriesZBDTE17, DBLP:conf/hicss/RodriguezPB19,  DBLP:conf/sigcse/BernuyZSPW21, DBLP:conf/lak/BernuyHSZLR0W22} as a bit more explicit intervention than mindless approaches.
For example, encouraging messages have been popularly employed to motivate students~\cite{DBLP:conf/lak/BernuyHSZLR0W22} or induce exercise behavior~\cite{DBLP:conf/ph/VriesZBDTE17}.
Furthermore, Liu \etal~\cite{DBLP:conf/cscw/LiuJPP14} proposed a method to motivate workers to return to the task of editing ill-formatted documents by designing feedback reflecting their progress based on the perspective of persuasive technology.
Their user study showed a significantly shorter average off-task time by adding visually encouraging feedback that reflected their progress.
Similarly, Rodriguez \etal~\cite{DBLP:conf/hicss/RodriguezPB19} suggested presenting context-aware content as a factor for successful nudge design, as it improves personal relevance and motivation.
Even though such nudging approaches can also be effective in our case, the inherent nature of nudging would not always be suitable in the context of procrastination on intellectually-demanding tasks.
Nudge has emphasized the importance of providing all available choices to people, without forbidding any of them~\cite{Thaler2008Nudge}, from the perspective of libertarian paternalism.
Meanwhile, people who procrastinate intentionally choose to do so while recognizing the necessity of completing the tasks~\cite{Lay1986Clarry, Frakes2016Procrastination}.
Given this, the effectiveness of such nudge-based approaches is uncertain for our situation, where the target tasks would require higher effort and focus than those in prior studies, such as editing ill-formatted documents.

Thus, to make these nudge-based approaches more sophisticated, we introduce an approach of providing the continuation of interrupted work to lower the hurdle for resuming a task.
As in the example of \name{} described in \secref{sec:intro}, workers prompted with imperfect but relevant work progress would try to work on it for 3--5 min instantly, which is important to avoid workers' perfectionism and procrastination behavior.
We leverage generative models for this purpose since recent large generative models are adept at producing context-aware content, although they are sometimes unable to satisfy workers who expect the generated content to replace parts of their tasks directly~\cite{DBLP:conf/iui/YangZZLL22, DBLP:conf/iui/KimMS21}.
This point guided us to anticipate that such generated content could be used to support workers' task engagement by drawing their diverted attention and offering a base to work on.
More specifically, we expected that workers can resume the interrupted task with less effort if they find the generated content supporting their ideation or providing relevant information.

\section{Proposed Approach: \proposed{}}
\label{sec:proposed}

In this section, we describe our approach of intervening with workers using the generated continuation of interrupted work.
We also elaborate on the strategy behind this idea from the perspective of persuasive technology.

\subsection{Description}
\label{sec:proposed-description}

\subsubsection{Overview}
\label{sec:proposed-description-overview}

\begin{figure*}[t]
    \centering
    \includegraphics[width=0.78\linewidth]{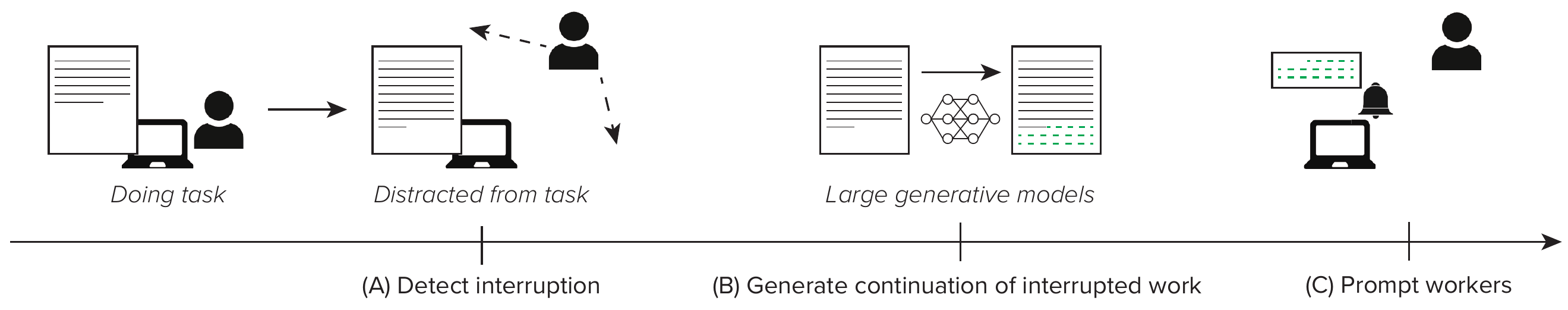}
    \caption{Overview of \proposed{}. (A) Detecting interruption based on workers' progress. (B) Generating continuation of interrupted work using large generative models. (C) Prompting workers to resume the task via notification.}
    \Description{This figure illustrates how workers can perform a task with the support of CatAlyst in a time series from left to right. In the leftmost part, it presents that a worker can be distracted while doing a task using icons of a user and computer used for the task. CatAlyst then detects the interruption (A) and generates a continuation of the interrupted work (B), as depicted using an icon of a large generative model. Then, it prompts the worker using the generated continuation (C), which is illustrated in the rightmost part using an icon of a bell that depicts the prompt from the computer.}
    \label{fig:proposed-overview}
\end{figure*}

\figref{fig:proposed-overview} shows an overview of our approach of intervening with workers to enhance task engagement.
\proposed{} first detects the moments where the progress of the workers' tasks is stopped from the interaction log (A).
It then generates content that is likely to follow interrupted work using large generative models (B).
For example, GPT-3~\cite{DBLP:conf/nips/BrownMRSKDNSSAA20} can be used in writing tasks to predict what would come next.
Finally, it sends notifications to the workers while showing part of the generated content to induce them to resume tasks (C).
To expand the applicability, we designed the pipeline so that we can use any generative model as long as it can produce a context-aware continuation of the interrupted task.
In this study, we implemented two prototypes: a writing support tool and a slide-editing support tool.
Both prototypes run as web applications.

\subsubsection{A: Detecting interruption}
\label{sec:proposed-description-detection}

Previous research has explored several approaches for detecting a break in a worker's concentration, including image-based sensing using a webcam, as in~\cite{DBLP:conf/chi/ArakawaY21}, and estimation from interaction log~\cite{DBLP:conf/cscw/LiuJPP14}.
In this study, we adopted the method of estimating using interaction logs to avoid the additional cost and risk of using sensors (\eg privacy).
Specifically, our prototypes send the interaction logs to a server and find a moment when the workers have not made any interaction for $T$~s.
The initial value of $T$ was determined in a pilot study as the approximate time that distinguishes the actual distraction of the workers from the random interval during an engagement, which was 45~s.
We were aware that, even if workers do not make any interaction over $T$~s, they might engage in tasks, such as long contemplation, resulting in false-positive detection.
Simultaneously, we presumed that intervening with them in such moments using the generated continuation would not work negatively, as it could provide them with new perspectives or information that could facilitate their contemplation.
This idea of designing an intervention that might work (or at least does not cause negative experiences) even when triggered by false-positive detection was inspired by Mindless Attractor~\cite{DBLP:conf/chi/ArakawaY21}.
Note that, as we received feedback from workers in our experiments, we can make the value $T$ controllable to provide them with global control of AI systems for a better human-AI relationship~\cite{DBLP:conf/chi/AmershiWVFNCSIB19}.

\subsubsection{B: Generating continuation of interrupted work}
\label{sec:proposed-description-generation}

\begin{figure*}[t]
    \centering
    \includegraphics[width=0.7\linewidth]{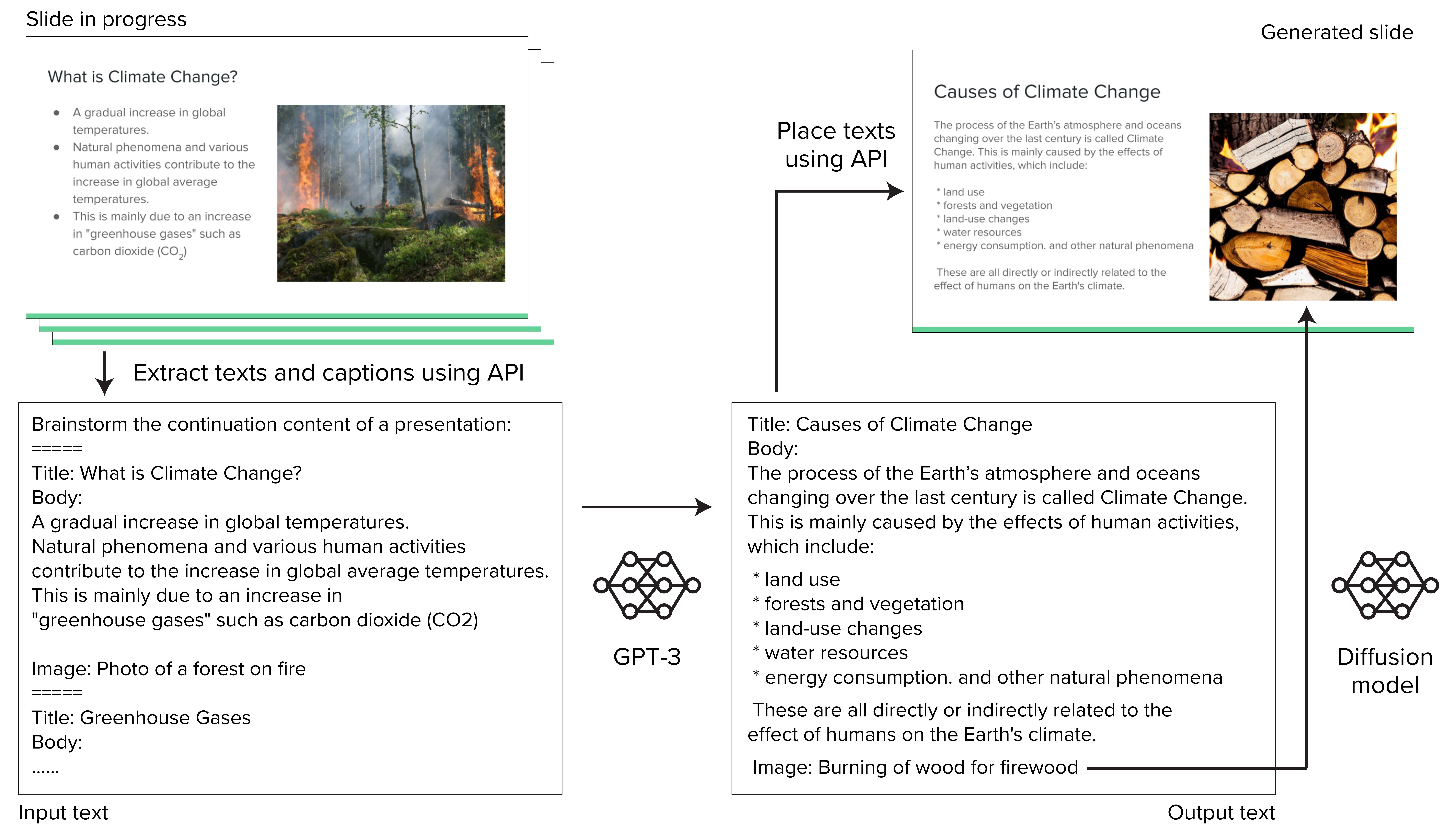}
    \caption{Actual example of how \proposed{} can generate the continuation in the slide-editing task using GPT-3.}
    \Description{This figure illustrates how CatAlyst generates the continuation in the slide-editing task. In the upper left part, screenshots of the slide in progress are presented, whose content (texts and the caption of photos) are extracted in the lower left part. The content is then provided to a GPT-3, and its output is presented in the lower right part. The output is subsequently used to construct a new slide, which is depicted in the upper right part.}
    \label{fig:proposed-slide}
\end{figure*}

We applied large generative models, such as GPT-3~\cite{DBLP:conf/nips/BrownMRSKDNSSAA20}, to generate the continuation of the interrupted work.
In the writing task, for instance, we provided the text in progress to a model for predicting the text to follow.
The same scheme can be applied to the slide-editing task; we extracted all text fragments and image captions from the slides in progress and provided them to a model (\figref{fig:proposed-slide}).
Subsequently, the continuation prediction from the model is used to fill the new slide that is automatically appended.
Here, \proposed{} can be applied to other domains by alternating the model to be used with, for example, those for music creation~\cite{DBLP:conf/chi/LouieCHTC20, DBLP:conf/chi/SuhYTC21} and drawing~\cite{DBLP:conf/chi/LinGCYY20, DBLP:conf/iui/KimMS21, DBLP:conf/acmidc/ZhangZWHSLHYY21, DBLP:conf/chi/ZhangYWLLYY22}.
Furthermore, the slide-editing task illustrates how our approach of harnessing large generative models can be extensively applied to new tasks by formulating appropriate inputs (\ie prompt programming~\cite{DBLP:conf/chi/ReynoldsM21}) even if those tasks have not been considered at the time of training.

\subsubsection{C: Prompting workers}
\label{sec:proposed-description-prompt}

Our prototypes notify workers using a Notification API of the browsers once the continuation is generated.
We aimed to draw not only workers' attention but also their interest, as discussed later in \secref{sec:proposed-strategy-attention}. 
Therefore, we designed the notification to present the beginning of the generated continuations (\eg the first sentence of the generated text or the title of the generated slide).
We note that \proposed{} can be extended to send notifications in other forms, such as using a smartphone to stimulate workers whose interest is deprived by smartphone apps.

\subsection{Strategy}
\label{sec:proposed-strategy}

\begin{figure*}[t]
    \centering
    \includegraphics[width=0.8\linewidth]{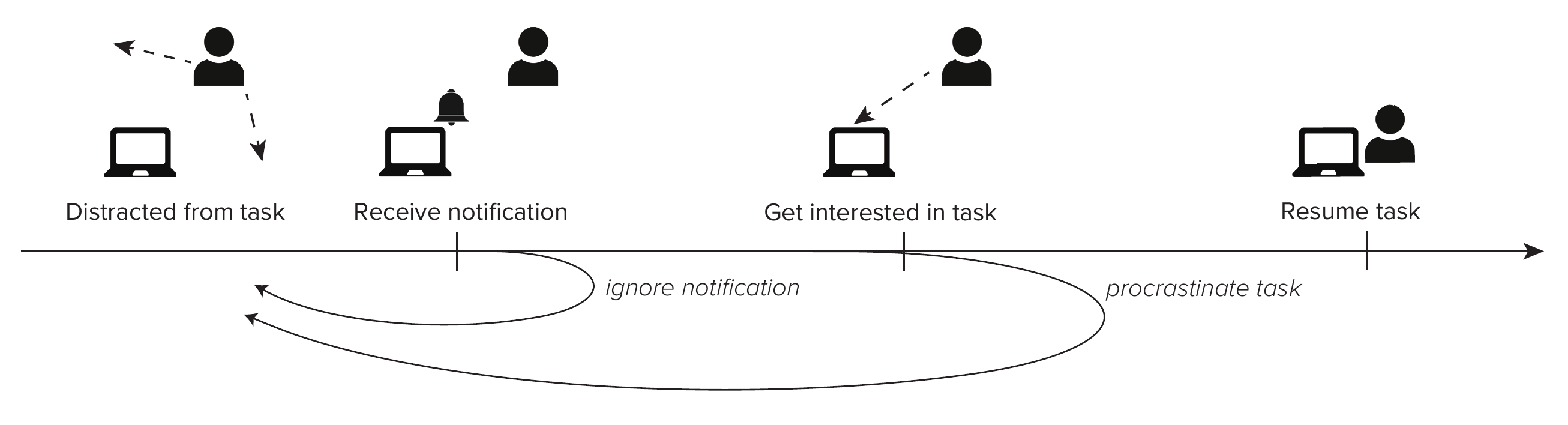}
    \caption{Interventions lead to workers' task resumption through multiple steps. At each step, there is a chance for an intervention to fail to induce workers' behavior change.}
    \Description{This figure illustrates the possible transitions that workers can experience during performing a task with the support of CatAlyst in a time series from left to right. In the leftmost part, the state that a worker is distracted is depicted using icons of a user and computer. The worker then receives a notification, but it is possible that they ignore the notification, which is illustrated using an arrow from the state to the initial state of being distracted. Even if the worker gets interested in the task, it is possible that they procrastinate the task, which is illustrated using an arrow from the state of having an interest in the task to the initial state of being distracted. In the rightmost part, the state that the worker resumes the task is presented.}
    \label{fig:proposed-funnel}
\end{figure*}

Resuming an intellectual task while a worker gets distracted is expected to require more cognitive load than in cases where a nudge or mindless intervention is successful, such as getting students' attention back~\cite{DBLP:conf/chi/ArakawaY21}.
In these situations, we expected that encouraging feedback, as in conventional methods~\cite{DBLP:conf/cscw/LiuJPP14, DBLP:conf/ph/VriesZBDTE17}, has room for improvement to provide more persuasive interventions and that prompting workers along with a continuation would be more effective.
By doing so, \proposed{} encourages workers to start thinking about the task regardless of their status, even for a short time, which is known as a popular practice to engage in the task~\cite{Pychyl2013Solving}.
This idea was originally inspired by a practitioner's insight, which has been popularly adopted by readers.
James Clear~\cite{James_Clear_2015} proposed a strategy for behavior change by employing an analogy of \textit{activation energy} in chemistry.
He explained that every habit has activation energy that is required to get started, and thus, it is important to introduce a \textit{catalyst} to lower the activation energy.
He further suggested utilizing intermediate steps with lower activation energy to induce behavior changes step by step.
Our strategy of achieving such a \textit{catalyst}-like effect using the generated continuation of interrupted work as the intervention can be broken down in detail, as shown in \figref{fig:proposed-funnel}.

\subsubsection{Guiding workers' interest to generated contents}
\label{sec:proposed-strategy-attention}

First, it is important that workers become interested in interventions and do not ignore them.
Regarding this point, interventions via notification can at least draw workers' attention but do not always draw their interest.
This means that it is easy to anticipate that workers will ignore them if the content of the intervention is not very useful~\cite{DBLP:conf/mhci/StreefkerkEN08} or attractive, as we often ignore push notifications on smartphones (the first failure case in \figref{fig:proposed-funnel}).
Here, Kovacs \etal~\cite{DBLP:journals/pacmhci/KovacsWB18} showed that adding variety to intervention is effective in maintaining novelty.
According to Berlyne~\cite{Berlyne1960Conflict}, inducing curiosity is important to drive motivation.
Wickens~\cite{DBLP:journals/ijhci/Wickens21} emphasizes the importance of inherent interest based on a computational model on top of cognitive science.

Our strategy is to make workers interested in the content of the intervention by presenting the generated continuation of their interrupted work.
For example, in the case of chatbots, people's curiosity about AI's outputs is reported as one of the major factors for them to use the technology~\cite{DBLP:conf/insci/BrandtzaegF17, Ling2021Factors}.
In this regard, \proposed{} generates continuation of interrupted work, which is more context-aware information, leading to the inherent interests of workers.

\subsubsection{Inducing workers' task resumption}
\label{sec:proposed-strategy-resumption}

The second part of the intervention flow (\figref{fig:proposed-funnel}) is to induce workers' behavior of resuming the task after they retrieve their interest in the original task.
More specifically, our objective is to prompt them to resume the task quickly for 3--5~min to avoid procrastination and engage in it.
According to Fogg's behavior model~\cite{DBLP:conf/persuasive/Fogg09a}, a behavior change is achieved when the elements of motivation, ability, and prompts are aligned.
Regarding motivation, if workers perceive \proposed{} as a collaborator, they will gain motivation to utilize the generated content, as demonstrated in prior studies on human-AI collaboration~\cite{DBLP:conf/acmidc/ZhangZWHSLHYY21, DBLP:conf/chi/ZhangYWLLYY22}.
Regarding ability, the generated continuation of the interrupted work can be a help, for example, by supporting their ideation or providing reference information, as in prior studies about human-AI co-creation reviewed in \secref{sec:rw-generative}.
Therefore, we expect \proposed{}'s design to guide workers to resume the original task with enhanced motivation and ability, preventing procrastination.

\subsection{Hypothesis}
\label{sec:proposed-hypothesis}

The design strategy of \proposed{}, as discussed thus far, poses the following hypotheses.
First, we should examine whether workers can recover their interest by being presented with the generated content as an intervention, as this is the first step toward resumption (see \figref{fig:proposed-funnel}).
\begin{quote}
    H1: \proposed{} is an effective means to keep attracting the interest of workers who are away from the task by presenting the continuation of interrupted work as an intervention.
\end{quote}

Next, even if the workers' interest returns to the task by seeing the generated continuation, they might not resume the task and go back to the distracted states, \ie procrastination, given the required hurdle for working on the task.
Therefore, we need to further investigate whether the workers prompted by the intervention change their behavior (\ie resume the task), guiding us to posit the following hypothesis.
\begin{quote}
    H2: \proposed{} can induce workers' behavior to resume the original task effectively through the intervention.
\end{quote}

If these hypotheses are supported, the effectiveness of \proposed{} in supporting workers' task engagement will be validated.
We further anticipate that the overall productivity of workers would be enhanced with the intervention, leading to the following hypothesis.
If this hypothesis is supported, \proposed{} can be used as an alternative approach to the previous studies to improve task efficiency.
\begin{quote}
    H3: \proposed{} can improve worker productivity by helping them avoid procrastination while performing tasks.
\end{quote}

Finally, we expect that \proposed{} can contribute to improving the subjective experience of workers who often face a high cognitive load during tasks.
If this hypothesis holds true along with the above hypotheses, we can conclude that our strategy behind the design of \proposed{} functioned as intended.
Therefore, we examine whether \proposed{} actually lowers the hurdle for workers to resume tasks and offers a favorable experience.
\begin{quote}
    H4: \proposed{} can lower the cognitive load imposed on workers while performing a task, thereby being favorably accepted by them.
\end{quote}

\section{Study 1: Writing Task}
\label{sec:study1}

To test our hypotheses, we first conducted a user study in a controlled setting in which we prepared a prototype of \proposed{} for supporting workers' writing.
This is because, while writing is one of the tasks that knowledge workers regularly perform, it requires a high level of skills and cognitive load, and workers often lose focus while performing it~\cite{Mastrangelo2006}.

\subsection{Design}
\label{sec:study1-design}

We conducted a within-participant study to compare three conditions: \textit{proposed}, \textit{control}, and \textit{none}.
Participants experienced interventions from our prototype of \proposed{} in the \textit{proposed} condition.
In the \textit{control} condition, they instead experienced a conventional intervention for supporting task engagement, that is, encouraging messages (see \secref{sec:rw-engagement}).
In the \textit{none} condition, they did not experience any intervention and only performed the writing task as usual.

\subsection{Task}
\label{sec:study1-task}

In this study, we had participants write essays of approximately 1,200 characters in Japanese.
We prepared six essay tasks based on publicized university exams to balance their difficulty, such as: ``There are more people who eat alone these days. What do you think about this trend from the perspective of modern social structure?''
Three were randomly chosen for each participant and assigned to the three conditions.
It took roughly 30~min to write one essay. 

\subsection{Implementation}
\label{sec:study1-implementation}

\begin{figure*}[t]
    \centering
    \includegraphics[width=0.9\linewidth]{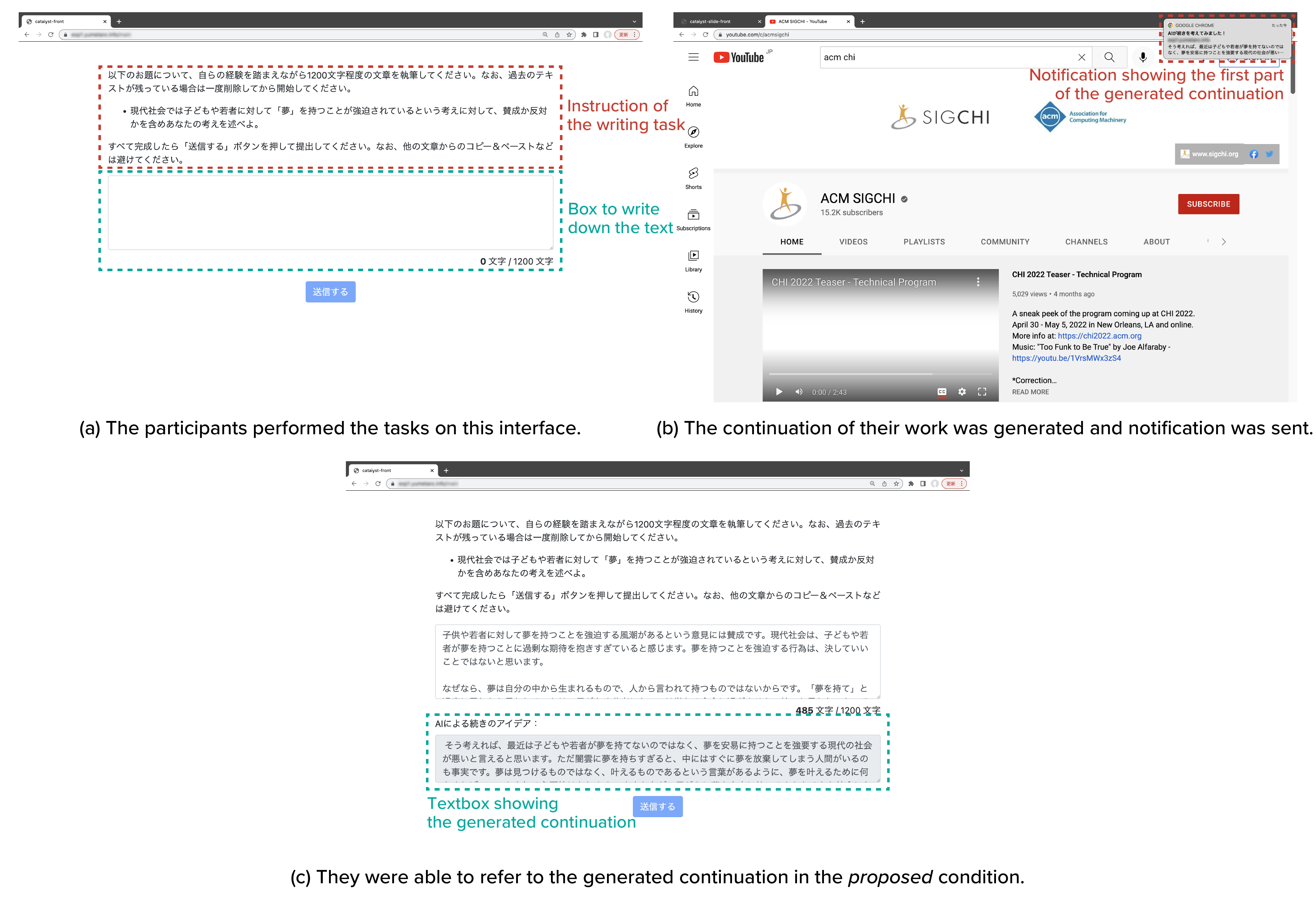}
    \caption{Interface used in Study 1.}
    \Description{This figure illustrates the interface of the prototype used for the study. In the upper left part, a screenshot of a web page presenting the instruction of the writing task at the top and a textbox to write down the text at the bottom is shown (A). In the upper right part, a screenshot of YouTube, which suggests that a user is distracted, and a notification showing the first part of the generated continuation at the top is shown (B). In the lower part, a screenshot of the same page as (A) is shown (C), but a textbox showing the generated continuation is added at the bottom.}
    \label{fig:interface-writing}
\end{figure*}

We prepared three types of web pages corresponding to the three conditions.
As shown in \figref{fig:interface-writing}, they had a center box in which the participants wrote texts.
In the \textit{proposed} and \textit{control} conditions, the web pages recorded the interaction log of the participants and detected the moment they got distracted.
In the \textit{proposed} condition, the written text in progress was sent to a remote server, where its continuation was generated using GPT-3~\cite{DBLP:conf/nips/BrownMRSKDNSSAA20} pretrained with the C4~\cite{DBLP:journals/jmlr/RaffelSRLNMZLL20} and CC100~\cite{DBLP:conf/acl/ConneauKGCWGGOZ20} datasets\footnote{\url{https://huggingface.co/rinna/japanese-gpt-1b}}.
The web page prompted the participants with a notification displaying its first sentence once the continuation was generated.
In the \textit{control} condition, the web page did not send the text and immediately showed a notification displaying an encouraging message that was randomly chosen from the six patterns prepared based on a public site blocker\footnote{\url{https://chrome.google.com/webstore/detail/blocksite-block-websites/eiimnmioipafcokbfikbljfdeojpcgbh}}.

\subsection{Procedure}
\label{sec:study1-procedure}

We published a web page for the experiment and recruited participants and, to secure the number of participants, we posted a link to the page on social media.
As a result, twelve participants (six males and six females; 24--59 years old) participated in the study.
Each two of them were assigned to one of the six possible orders of the three conditions. 
The participants agreed to the research policy and were provided with a brief explanation of the usage of the web pages in the beginning.
Here, we clarified that we do not penalize them even if they spend a long time on the tasks so that they can perform the tasks as usual.
In addition, we told them that they were allowed to perform any habit they might have in performing tasks, for example, listening to music, to induce spontaneous behavior, following the procedure of Liu \etal~\cite{DBLP:conf/cscw/LiuJPP14}.
They then performed three essay tasks using the web page that corresponded to the assigned order.
Every time they completed a task, they were asked to fill out a questionnaire to evaluate their experience of the task using the web page.
They were also asked to fill out another questionnaire that requested a comparison of the three conditions and additional comments after completing all the assigned tasks.
The study was conducted remotely for all participants.

\subsection{Measure}
\label{sec:study1-measure}

We used multiple measures according to our hypotheses.
From the participants' behavior during the use of web pages, we calculated four measures: interest retrieval time, ignorance rate, progress after resumption, and total writing time.
From the questionnaire responses, we evaluated cognitive load and system usability.

\subsubsection{Interest retrieval time}

In correspondence with H1, we calculated \textit{interest retrieval time}, the time each participant spent from the moment they received a notification to the moment they recovered their interest in the task.
Here, we caught the recovery of interest by finding the moment when they made any operations on the web page (\eg moving a mouse or pressing a key) for the first time after receiving a notification.
Furthermore, we compared the values of the \textit{proposed} and \textit{control} conditions because the \textit{none} condition did not present any notifications.
When the interest retrieval time of the \textit{proposed} condition was shorter than that of the \textit{control} condition, H1 is supported.

\subsubsection{Ignorance rate}

We also calculated \textit{ignorance rate}, a rate of interventions (\ie notifications) that were ignored by the participants.
The ignorance rate is highly correlated with the interest retrieval time (\ie a longer interest retrieval time leads to a higher ignorance rate) because the \textit{proposed} and \textit{control} conditions prompted a notification every $T$~s until the participants made any operations.
Still, we prepared this measure to evaluate H1 precisely.
Specifically, as discussed in \secref{sec:rw-generative}, there was a risk of losing the trust of the participants due to a mismatch with their expectations, fostering ignorance of the notification.
Therefore, to confirm that \proposed{} can keep attracting their interest, we examined how the ignorance rate changed as the number of interventions each participant received increased.
If the ignorance rate of the \textit{proposed} condition was lower than that of the \textit{control} condition and exhibited stable values with respect to the increase in the number of interventions, H1 is further supported.

\subsubsection{Progress after resumption}

In correspondence with H2, we quantified and compared the number of characters the participants typed within a specific period after the participants recovered their interest in the task, namely, \textit{progress after resumption}.
As discussed in \secref{sec:proposed-strategy-resumption}, they might procrastinate on the task even though they have recovered their interest.
In such cases, progress after resumption would remain insignificant.
Conversely, H2 is supported if the increase in the number of characters after resumption in the \textit{proposed} condition was more significant than that in the \textit{control} condition.

\subsubsection{Total writing time}

We evaluated the productivity of the participants via two measures.
The first is the \textit{total writing time} they spent on the tasks.
The average of the total writing time would be equivalent if \proposed{} did not contribute to the improvement of worker productivity since we balanced the difficulty and assignments of the tasks prepared.
In other words, H3 is not supported unless the total writing time of the \textit{proposed} condition was shorter than that of the other two conditions.

\subsubsection{Subjective quality}

We also evaluated the \textit{subjective quality} of the texts the participants wrote.
This was to ensure that the reduction in the total writing time was enabled not by sacrificing the text quality, but by the improvement in their productivity.
Therefore, we asked three volunteer raters who had not participated in the experiment to rate the texts in a manner agnostic to the experimental conditions.
Each text was evaluated for three aspects using a 7-point Likert scale: consistency, readability, and overall quality.
Therefore, H3 is supported when the \textit{proposed} condition shortened the total writing time while the raters' evaluation for the texts written in this condition, at least, did not deteriorate compared to the other two conditions. 

\subsubsection{Cognitive load}

We then examined H4 using the questionnaire.
Specifically, we used the six items of the NASA-TLX~\cite{Hart1988} and calculated the raw TLX score~\cite{Byers1989}.
This allowed us to examine \textit{cognitive load} across six factors: mental demand, physical demand, temporal demand, performance, effort, and frustration.
As higher scores indicate higher cognitive load, H4 is supported when the scores of the \textit{proposed} condition were lower than those of the other conditions.

\subsubsection{System usability}

Furthermore, we examined \textit{system usability} to complement the cognitive load in terms of evaluating the subjective experience of the participants.
We adopted ten items from the System Usability Scale (SUS)~\cite{Brooke1996} and compared their usability evaluation for the \textit{proposed} and \textit{control} conditions.
The case in which the score of the \textit{proposed} condition was higher than that of the other conditions corroborates H4 because it implies better usability of \proposed{}.
Moreover, we asked the participants to rank the three conditions in order of their evaluation of system usability in the questionnaire they completed at the end of the study.
We compared their responses across the three conditions to confirm that the \textit{proposed} condition was ranked higher than the other conditions, which leads to supporting H4.

\subsection{Results}
\label{sec:study1-results}

\subsubsection{Interest retrieval time}

First, we calculated the interest retrieval time, as presented in \tabref{tbl:study1-interest}.
The participants in the \textit{proposed} condition reacted to the intervention within a much shorter period than those in the \textit{control} condition.
Welch's $t$-test after Levene's test confirmed that the mean of the values in the \textit{proposed} condition was significantly shorter ($t(72.31) = 3.20, p = 0.002$) than that in the \textit{control} condition, supporting H1.
Furthermore, the \textit{control} condition exhibited a greater variance in the interest retrieval time.
In fact, we found cases where the encouraging message did not work for some completely distracted participants, whose interests were recovered after more than 10 min, conforming to previous research \cite{DBLP:conf/chi/MarkGK08}.

\subsubsection{Ignorance rate}

Moreover, H1 was corroborated by a comparison of ignorance rates.
Corresponding to the longer interest retrieval time in the \textit{control} condition, it presented a much higher number of notifications until the participants recovered their interest in the task, most of which were ignored, as shown in \tabref{tbl:study1-ignorance} (A).
The ignorance rate of the \textit{proposed} condition was significantly lower ($p < 0.001$) than that of the \textit{control} condition, according to Fisher's exact test.
Interestingly, we could not find a difference between the two conditions when we focused only on the first intervention presented to each participant ($p = 0.260$).
In other words, the intervention of the \textit{control} condition attracted the participants' interest at the beginning; however, as they saw more encouraging messages, they became less persuaded by the encouraging messages, leading to an increase in the ignorance rate.
Conversely, this suggests that \proposed{} can keep attracting the interest of workers, which we intended by introducing the personalized, context-aware, and variational intervention.

In addition, this result can be explained by \textit{perceived utility}, which Ling \etal~\cite{Ling2021Factors} mentioned as a factor in maintaining the use of AIs.
The workers will keep interested in the intervention if they feel usefulness in the generated content (\eg as support for ideation), whereas conventional encouraging interventions do not have such utility.
We later discuss this point based on the \textit{system usability} and comments of the participants.

\begin{table}[t]
    \centering
    \caption{Comparison of the interest retrieval time between the \textit{proposed} and \textit{control} conditions, which were significantly different ($p = 0.002$).}
    \label{tbl:study1-interest}
    \begin{tabular}{cr@{\hspace{0.5em}}c@{\hspace{0.3em}}r}
        \toprule
                 & mean  &       & \multicolumn{1}{@{\hspace{-0.5em}}c@{\hspace{0.6em}}}{SD} \\
        \midrule
        control  & 135.0 & $\pm$ & 303.7 (s) \\
        proposed & 18.8  & $\pm$ & 25.1  (s) \\
        \bottomrule
    \end{tabular}
\end{table}

\begin{table}[t]
    \centering
    \caption{Comparison of the ignorance rate between the \textit{proposed} and \textit{control} conditions. While the rate was significantly lower for the {proposed} condition ($p < 0.001$), no difference was observed when we focused on only the first intervention presented to each participant ($p = 0.260)$.\protect\footnotemark}
    \label{tbl:study1-ignorance}
    \begin{tabular}{crrrr}
        \toprule
                 & \multicolumn{2}{c}{(A) \textbf{All}} & \multicolumn{2}{c}{(B) \textbf{First time}} \\
        \cmidrule(lr){2-3} \cmidrule(lr){4-5}
                 & ignored & worked                     & ignored & worked                            \\
        \midrule
        control  & 185     & 71                         & 4       & 11                                \\
        proposed & 5       & 29                         & 0       & 8                                 \\
        \bottomrule
    \end{tabular}
\end{table}

\subsubsection{Progress after resumption}

We\footnotetext{Total number of the notification presented to the participants as the first intervention in the \textit{control} condition exceeds the number of participants because some of them did not recover their interest until several notifications were prompted even in the first intervention.}
 then examined the progress after resumption, as presented in \tabref{tbl:study1-progress}.
We found that the participants in the \textit{proposed} condition made more progress within $T$~s after their interest returned to the task than those in the \textit{control} condition, according to the $t$-test ($t(98) = 2.08, p = 0.041$).
This implied two possibilities regarding the participants' behavior: 1) participants in the \textit{proposed} condition could utilize the generated content to come up with what to write next and 2) participants in the \textit{control} condition procrastinated the task even after being guided to perform some operation on the web page by encouraging messages.
This result supported H2.

\subsubsection{Total writing time}

As shown in \tabref{tbl:study1-time}, we could not find a significant difference in the total time the participants spent between the three conditions, according to the Friedman test after Levene's test.
Thus, H3 was not supported.
However, we can see that the participants in the \textit{proposed} condition spent less time on average than those in the other conditions, which implies the effect of \proposed{}.
Simultaneously, the average time spent in the \textit{control} condition was slightly longer than that in the \textit{none} condition.
This can be attributed to the fact that the previous work in \secref{sec:rw-engagement} employed a clear criterion for detecting interruptions, such as being presented with explicit interruption cues by an experimenter~\cite{DBLP:conf/cscw/LiuJPP14}, whereas we used time-based na\"{i}ve criteria to examine the effectiveness of \proposed{} against false-positive detection.

\begin{table}[t]
    \centering
    \caption{Comparison of the progress after resumption time between the \textit{proposed} and \textit{control} conditions, which were significantly different ($p = 0.041$).}
    \label{tbl:study1-progress}
    \begin{tabular}{cr@{\hspace{0.5em}}c@{\hspace{0.3em}}r}
        \toprule
                 & mean  &       & \multicolumn{1}{@{\hspace{-0.5em}}c@{\hspace{0.6em}}}{SD} \\
        \midrule
        control  & 17.7  & $\pm$ & 36.3 (chars)  \\
        proposed & 63.5  & $\pm$ & 186.2 (chars) \\
        \bottomrule
    \end{tabular}
\end{table}

\begin{table}[t]
    \centering
    \caption{Comparison of the total writing time between the three conditions, which were not significantly different ($p = 0.174$).}
    \label{tbl:study1-time}
    \begin{tabular}{cr@{\hspace{0.5em}}c@{\hspace{0.3em}}r}
        \toprule
                 & mean  &       & \multicolumn{1}{@{\hspace{-0.5em}}c@{\hspace{0.6em}}}{SD} \\
        \midrule
        control  & 2557.3 & $\pm$ & 2196.5 (s) \\
        proposed & 1747.3 & $\pm$ & 1660.0 (s) \\
        none     & 2231.2 & $\pm$ & 1359.8 (s) \\
        \bottomrule
    \end{tabular}
\end{table}

\subsubsection{Subjective quality}

Meanwhile, we also could not find a significant difference in the raters' subjective evaluation of the quality of the texts the participants wrote between the three conditions, according to the Friedman test.
Specifically, as shown in \tabref{tbl:study1-quality}, all of the three aspects the raters evaluated (\ie consistency, readability, and overall quality) did not exhibit a significant difference ($p = 0.354$, $0.221$, and $0.972$, respectively).
This result implied that \proposed{} did not deteriorate the quality of the texts the participants wrote, while H3 was not supported.

\begin{table}[t]
    \centering
    \caption{Comparison of the subjective quality of the texts between the three conditions, which were not significantly different regarding consistency, readability, and overall quality ($p = 0.354$, $0.221$, and $0.972$, respectively).}
    \label{tbl:study1-quality}
    \begin{tabular}{cccc}
        \toprule
                 & consistency     & readability     & overall quality    \\
        \midrule
        control  & 5.50 $\pm$ 1.21 & 5.25 $\pm$ 1.32 & 5.11 $\pm$ 1.37  \\
        proposed & 5.03 $\pm$ 1.61 & 5.17 $\pm$ 1.63 & 4.89 $\pm$ 1.65  \\
        none     & 5.47 $\pm$ 1.54 & 4.97 $\pm$ 1.46 & 5.06 $\pm$ 1.43  \\
        \bottomrule
    \end{tabular}
\end{table}

\begin{table}[t]
    \centering
    \caption{Comparison of the participants' evaluations of the system usability between the \textit{proposed} and \textit{control} conditions, which were not significantly different ($p = 0.082$).}
    \label{tbl:study1-sus}
    \begin{tabular}{cr@{\hspace{0.5em}}c@{\hspace{0.3em}}c@{\hspace{0.8em}}}
        \toprule
                 & mean  &       & SD   \\
        \midrule
        control  & 62.1  & $\pm$ & 22.8 \\
        proposed & 75.6  & $\pm$ & 13.7 \\
        \bottomrule
    \end{tabular}
\end{table}

\subsubsection{Cognitive load}

\begin{figure*}[t]
    \centering
    \includegraphics[width=0.85\linewidth]{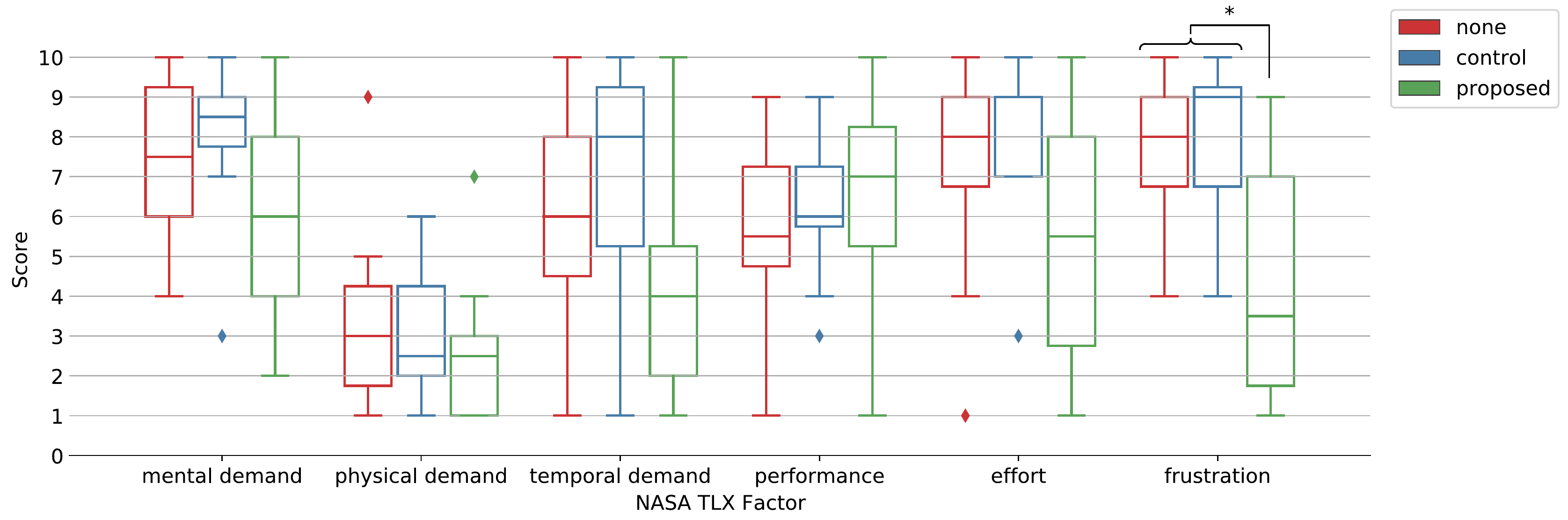}
    \caption{Participants' evaluations of their cognitive load in Study 1. The frustration they felt in the \textit{proposed} condition was significantly lower than those of the \textit{control} ($p = 0.019$) and \textit{none} ($p = 0.019$) conditions.}
    \Description{This chart shows the result of the cognitive load measured by the NASA-TLX test. There are six factors from left to right: mental demand, physical demand, temporal demand, performance, effort, and frustration. Each with three box plots corresponds to the none, control, and proposed condition. The y-axis shows the score spanning from 0 to 10. For most factors, the proposed condition exhibits a trend of lower scores compared with the other conditions, implying a lower cognitive load. In addition, for the frustration factor, a star illustrating the existence of a significant difference in the scores between the proposed condition and the others is shown.}
    \label{fig:study1-tlx}
\end{figure*}

To evaluate H4, we also examined the participants' evaluations of their cognitive load, as shown in \figref{fig:study1-tlx}.
One-way repeated measures ANOVA suggested significant differences in their cognitive load with respect to the frustration factor ($F(2, 22) = 9.37, p = 0.001$).
We then found that the frustration they felt during the \textit{proposed} condition was significantly lower than those of the \textit{control} ($t(11) = 3.36, p = 0.019$) and \textit{none} ($t(11) = 3.25, p = 0.019$) conditions, according to the post-hoc test with Holm correction.
This result suggests the plausibility of H4.

\subsubsection{System usability}

Finally, we examined the participants' evaluations of the system usability.
As presented in \tabref{tbl:study1-sus}, the scores obtained using SUS implied that the participants preferred the \textit{proposed} condition over the \textit{control} condition, while their difference was not significant according to the paired $t$-test ($t(11) = 1.91, p = 0.082$).
At the same time, the participants' rankings regarding usability suggested that the three conditions made a significant difference in their rank order, according to the Friedman test ($\chi^2(2) = 8.91, p = 0.012$).
Specifically, the post-hoc Nemenyi test~\cite{nemenyi1963distribution} revealed that the \textit{proposed} condition was likely to be ranked significantly higher (\ie regarded as having better usability) than the \textit{none} condition ($p = 0.022$).
In particular, we found that eight participants responded that the \textit{proposed} condition was the best.
These results indicated that \proposed{} was favorably accepted by participants, corroborating H4.
This also implied that the \textit{proposed} condition contributed to the perceived utility of the participants, and thus, its effectiveness did not deteriorate after repeated exposure, as mentioned above.

\subsection{User Comments}
\label{sec:study1-comments}

From the comments the participants left, we can confirm that \proposed{} functioned as we expected, as follows.
\begin{quote}
    Not only did the notification make me realize that I was distracted, but it also helped me get back to the task as I wondered what the AI had written.
\end{quote}
\begin{quote}
    I often lost concentration and stopped working because I could not come up with the next sentence, so it was easier to get back to work with being presented with the AI's text.
\end{quote}

Furthermore, we found that the generated continuation helped their work by providing ideas and relevant information.
\begin{quote}
    I was very surprised the first time when the text from the AI was presented. It actually fit the topic, and I was able to use that as a reference for my writing.
\end{quote}
\begin{quote}
    Although I did not find consistency or uniformity in the AI's output, I found it useful for imitating its writing style and obtaining ideas for paraphrasing.
\end{quote}
The latter comment suggested that the participant positively perceived the use of \proposed{} even though the performance of the large language model was not completely satisfactory.
Importantly, such a gap in performance expectations can lead to devastating trust, as discussed in \secref{sec:rw-generative}.
However, our design reframes the main goal of \proposed{} to guide participants' interest instead of directly contributing to their tasks.
We suspect that this induced the participants' tolerance of the AI's performance and the enhancement of their subjective usability.

Our observations are further supported by examining their comments on the \textit{control} condition, as follows:
\begin{quote}
    At first, it was a good trigger for recognizing that I was not concentrating. But, when I received the notification from the system while I interrupted the task to write a quick reply to an incoming message, I felt a little annoyed.
\end{quote}
This comment not only emphasizes the importance of designing an intervention that also works with false-positive detection but also suggests the reason why the intervention of the control condition tended to be ignored as the number of interventions increased (\tabref{tbl:study1-ignorance}).

We note that one participant requested that the frequency of the notification be controllable.
This conforms to the previous study that emphasized the importance of providing global control in AI-based systems~\cite{DBLP:conf/chi/AmershiWVFNCSIB19}.
This comment helped us improve the prototype we provided in the following long-term study.

\section{Study 2: Writing Task in the Wild}
\label{sec:study2}

Thus far, the effectiveness of \proposed{}'s design in the writing task has been validated by confirming H1, H2, and H4 through the controlled user study.
Still, the results may be attributed to the fact that it might be the first time that the participants interacted with the recent large generative models.
In other words, conducting a long-term study to deny the possibility that the effectiveness of \proposed{} depends on freshness is desirable.
Therefore, we evaluated the effectiveness of \proposed{} in a study with a longer duration and participants having diverse motivations and purposes for writing.
Ten participants used the prototype for five days and provided their perspectives and feedback on the design of \proposed{}.

\subsection{Implementation}
\label{sec:study2-implementation}

We reused the prototype used in \secref{sec:study1}.
We enabled participants to adjust the time threshold $T$, which would be helpful, particularly when the prototype is used in the wild, based on the feedback we had gained in the previous study.
Furthermore, we assumed that participants would leave the web page for a substantial amount of time because of other tasks in their daily lives (\eg having lunch) in contrast to the controlled setting in the first study.
We thus added a feature of sending participants an email with generated content 5~min after they left the web page.
We expected that this feature would help participants to resume the original task when they finished other tasks.
Note that the content written by participants was neither accessible from experimenters nor stored on the server to preserve their privacy; it was only used for AI's inference and immediately discarded after it.

\subsection{Procedure}
\label{sec:study2-procedure}

\begin{table}[t]
  \centering
  \caption{Backgrounds of the participants in Study 2.}
  \label{tbl:participants-stats}
  \begin{tabular}{rcc@{\hspace{1.5em}}l}
    \toprule
        & \multirow{2}{*}{Gender} & Writing & \multirow{2}{*}{Main usage} \\
        &                         & habit       & \\
    \midrule
    P1  & M      & 3 years              & Novel                       \\
    P2  & F      & 8 years              & Web article / Personal blog \\
    P3  & F      & 2 years              & Diary                       \\
    P4  & M      & 0.5 years            & Personal blog               \\
    P5  & F      & 4 years              & Web article / Personal blog \\
    P6  & F      & rarely               & University report           \\
    P7  & M      & 10 years             & Diary                       \\
    P8  & M      & 1 year               & Fantasy / Lyric writing     \\
    P9  & M      & rarely               & Private letter              \\
    P10 & M      & 6 years              & Novel                       \\
    \bottomrule
  \end{tabular}
\end{table}

In the same manner as in Study 1, ten participants (P1--P10) were recruited.
\tabref{tbl:participants-stats} summarizes their backgrounds.
They self-reported their writing experience regarding the period they regularly had writing tasks.
In this study, the participants first agreed to the research policy and familiarized themselves with how to use the prototype.
We asked them to use the prototype for writing freely during the study period whenever they would like, without setting any quota.
After a participant spent five days, we asked for their comments to understand how they perceived the use of \proposed{} in terms of its usability and influence on their behavior.
The participants were asked a series of questions: ``\textit{what kind of writing did you use the system for?}''; ``\textit{can you tell us your overall impressions of using the system?}''; ``\textit{did you feel any change with and without the system?}''; ``\textit{what is your impression about the AI which generated content?}''; ``\textit{what was this system like for you?}''; ``\textit{what would you like to see with this system in the future?}''; ``\textit{do you want to continue using the system?}''
Note that we conducted the experiment remotely for all participants.

After the above procedure, we also examined the participants' use of \proposed{} from the interaction log.
This aimed to confirm the long-term effectiveness of \proposed{} in a manner disassociated from the possible effect of the freshness in interacting with large generative models.
Specifically, we calculated the \textit{interest retrieval time} in the same manner as in the previous study and analyzed its trend over the five days of their use.

\subsection{Usage results}
\label{sec:study2-usage}

\begin{table}[t]
    \centering
    \caption{Comparison of the interest retrieval time over the five days of the participants' use, which were not significantly different ($p = 0.791$).}
    \label{tbl:study2-usage}
    \begin{tabular}{lrrrrr}
        \toprule
        day                              & 1    & 2     & 3     & 4     & 5    \\
        \midrule
        \# of detected interruptions & 31   & 49    & 10    & 13    & 14   \\
        average retrieval time           & 95.4 & 117.1 & 121.1 & 105.2 & 93.4 \\
        standard deviation               & 81.9 &  91.4 & 102.7 &  95.4 & 89.6 \\
        \bottomrule
    \end{tabular}
\end{table}

As presented in \tabref{tbl:study2-usage}, we could not find a significant difference in the interest retrieval time across the five days, according to one-way ANOVA after Levene's test ($F\left(4, 112 \right) = 0.425, p = 0.791$).
This suggested that the effectiveness of \proposed{} in guiding the participants' interests did not disappear even after long exposure to it.
We note that the higher values in \tabref{tbl:study2-usage} compared to the previous experiment (\tabref{tbl:study1-interest}) are attributed to the difference in the situations.
Specifically, these values include the moments when the participants interrupted their writing tasks for necessary reasons in their daily lives, while we asked the participants in \secref{sec:study1} to complete the assigned tasks as a part of the experiment.

In addition, \tabref{tbl:study2-usage} allows us to infer that the participants continued the use of \proposed{} over the five days, whereas the occasions they used can be influenced by various factors (\eg weekdays or weekends) in the uncontrolled setting of this experiment.
Note, since we did not store the content written by the participants, the progress they made cannot be quantified.
Therefore, we then analyzed their responses in the semi-structured interviews we conducted.

\subsection{Interview Results}
\label{sec:study2-results}

Overall, participants responded positively to the question, ``\textit{can you tell us your overall impressions of using the system?}'', and mentioned various benefits of the prototype.
Moreover, eight (all participants except P7 and P10) showed their motivation for continued use toward the question: ``\textit{do you want to continue using the system?}''
This result corroborated the support of H4, that is, the use of \proposed{} offers a favorable experience, and workers feel its usefulness.
Moreover, four of them (P4, P5, P6, and P9) mentioned the novelty of the outputs as a key reason for their willingness to use the prototype.
\begin{quote}
    I've always been curious about what's generated. They were interesting. The excitement never stopped during the use and I'd not get bored. (P6)
\end{quote}
These comments implied that the participants' favorable responses in \secref{sec:study1} are not merely due to its freshness but thanks to the design of \proposed{}.
Specifically, the unlimited variety in the outputs of the generative models plays a key role in keeping attracting workers' interest in the intervention, as we discussed in \secref{sec:proposed-strategy-attention}.
The implications we obtained from the responses to the other questions are summarized below.

\subsubsection{Effects on behavior}

Seven participants (P2, P3, P4, P5, P6, P8, and P9) mentioned a significant change in their behavior owing to the system as an answer to the question: ``\textit{did you feel any change with and without the system?}''
\begin{quote}
    The notifications helped me cut out the time I spent touching my smartphone. I also sometimes started or proceeded to write sentences because I was curious to see what kind of sentences the AI would generate. I found this to be effective in the initial stages, when the writing was the heaviest. (P3)
\end{quote}
The comments were parallel to those in the short study (\secref{sec:study1-comments}) and demonstrated how \proposed{} affects workers and prevents their task procrastination.
In particular, P3's comment implied a \textit{catalyst}-like effect by lowering the hurdle for resuming the task, supporting our strategy.

\subsubsection{Feelings about AI's accuracy}
\label{sec:study2-results-accuracy}

Next, the participants responded similarly to the question: ``\textit{what is your impression about the AI which generated content?}''
More specifically, they pointed out the variance in the quality of the generated content.
\begin{quote}
    For my creative writing, the tool provided me with a good stimulus and motivation to continue writing because it allowed me to choose plots and words that I could not have reached on my own. However, I could not directly leverage it well for articulating my thought, like in a diary. (P8)
\end{quote}
\begin{quote}
    I felt that AI has its strengths and weaknesses depending on the field of articles I write. As for abstract writing for hobbies, such as travelogues, novels, diaries, \textit{etc}., the AI helped me think about how to phrase words and what to do next, and it worked beyond my expectations. However, AI generation was not appropriate when it requires a great deal of checking, like using it for my job [web writing]. (P2)
\end{quote}
\begin{quote}
    I was trying to write a full-length novel game scenario, and there was a constraint that each character setting was already fixed. While the outputs were interesting and I had fun seeing them, I needed to manually edit them, such as wording in generated utterances. (P10)
\end{quote}
This result highlighted the difficulty of expecting a large generative model to perform well in all domains without any tuning, as we discussed in \secref{sec:rw-generative}.
However, it is noteworthy that the participants affirmatively perceived their experiences of using \proposed{} owing to its design of not aiming to directly replace part of their tasks in the first place.

\subsubsection{Role of \proposed{}}

The participants' responses shed light on the three different roles of \proposed{} regarding the question: ``\textit{what was this system like for you?}''

\vspace{.5em}\noindent\textbf{\proposed{} as a reminder.} Two participants (P2 and P9) appreciated the effective reminder function.
\begin{quote}
    I used the system for my job. Due to the working-from-home situation, I often get distracted. I appreciate the reminding function as well as its generated content, which lowers the hurdle for writing. (P2)
\end{quote}

\noindent\textbf{\proposed{} as an ideator.} The benefit of obtaining ideas from \proposed{} was also emphasized by some participants (P1, P4, P8, and P10), who mostly performed creative writing, as follows:
\begin{quote}
    The suggested wording and story flow can be enough hints. For example, I got inspired to include a monologue of the protagonist or conversation to improve the tempo of the story. I used them to accelerate my writing. (P1)
\end{quote}
This benefit is aligned with previous works supporting writing with generative models and confirms the validity of our expectation that \proposed{} can induce task resumption in workers by improving their ability to perform the task (\secref{sec:proposed-strategy-resumption}).

\vspace{.5em}\noindent\textbf{\proposed{} as a peer.} Other participants (P2, P3, P5, and P6) perceived \proposed{} as a peer or collaborator who motivated them:
\begin{quote}
    I found it like a co-worker working with me. I became curious about what they produce while I was taking a break. (P5)
\end{quote}
\begin{quote}
    It is easier to expand my thinking by reviewing AI's outputs rather than contemplating from scratch by myself. In that sense, the AI was a peer to whom I can talk. (P3)
\end{quote}
Their responses corresponded to our expectation that \proposed{} can gain workers' interest (\secref{sec:proposed-strategy-attention}) by stimulating their curiosity and can induce their task resumption by increasing their motivation (\secref{sec:proposed-strategy-resumption}).

\subsubsection{Room for further improvement of \proposed{}}
\label{sec:study2-results-improvement}

The participants also mentioned their requests for the system when answering the question: ``\textit{what would you like to see with this system in the future?}''
Some participants (P1, P4, and P7) mentioned the need to control notification timing as follows:
\begin{quote}
    Ideally, I want to turn off the notification when I'm deeply thinking about the story flow, even if I'm not typing any words. (P1)
\end{quote}
There is room for improving its accuracy because our detection approach is a practical but na\"{i}ve method.
\begin{quote}
    I hope the system cooperates with a smartwatch because I think it would be useful to receive notifications when I am walking or something like that, so I can think about it. (P4)
\end{quote}
This comment implied a potential need for workers to actively set an intervention timing for \proposed{} as a reminder while performing different tasks (\eg walking).
Moreover, there is an interesting conflict in the comments.
\begin{quote}
    I could not wait for the AI's generation. It would have been even better if there had been a button for actively using it or more suggestions from the system. (P7)
\end{quote}
\begin{quote}
    I like its current design. If there had been an auto-generate button, it would have been a bad experience if the output had been inaccurate. In the current case, by thinking of it as an added feature of a reminder, my expectations were relaxed and I was glad to see it for the fun of it. (P6)
\end{quote}
In \secref{sec:disc-passive}, we further discuss this difference in the participants' needs and why we did not implement such a button for actively using generative models based on these comments.

Three participants (P1, P8, and P10) wanted to see multiple generations simultaneously.
\begin{quote}
    I would appreciate it if it could suggest several patterns of sentences at the same time so that I can do a lot of thinking when I receive the notification. (P1)
\end{quote}
This demand can be explained by prior research suggesting that presenting multiple perspectives, including AI's suggestions, can foster one's reflection by offering discussion grounds~\cite{DBLP:conf/chi/ArakawaY20}.

Additionally, two participants (P6 and P10) requested \proposed{} implemented in existing editors.
\begin{quote}
    It's interesting, but I'd prefer to keep using my current editor [Google Docs] simply because it has more useful functions. If there is an API for integrating this system into my editor, I'd appreciate it. (P10)
\end{quote}
In fact, \proposed{} can be easily integrated into online editors, as we will show a prototype in the form of a Chrome extension for supporting slide-editing on Google Slides in the next study.

\section{Study 3: Slide-Editing Task}
\label{sec:study3}

The short- and long-term studies in Sections~\ref{sec:study1} and \ref{sec:study2} demonstrated the effectiveness of \proposed{} in writing tasks.
We then conducted another study in which participants performed slide-editing tasks with the support of \proposed{}.
This study aimed to examine the domain extensibility of our design, particularly when the large generative model used in \proposed{} was not tuned for this specific task.
We again evaluated the hypotheses in \secref{sec:proposed-hypothesis} in a similar manner as in \secref{sec:study1}, a within-participant design involving the \textit{proposed}, \textit{control}, and \textit{none} conditions.

\subsection{Task}
\label{sec:study3-task}

We prepared three tasks in which participants were asked to create a slide of at least 15 pages about an aging society, online abuse, or global warming.
Here, we prepared a six-page draft slide for each topic and provided it to start the corresponding task.
This aimed to balance the difficulty of the tasks because, without the draft, the quality of the slides and the time required to edit them would largely depend on each participant.
It took roughly 30~min to complete a single task.
We randomly assigned the three tasks to three conditions for each participant.

\subsection{Implementation}
\label{sec:study3-implementation}

In this study, we asked participants to use Google Slides\footnote{\url{https://www.google.com/slides/about}} to edit the slides after installing a Chrome extension that we developed.
The extension detects the moment when the participants become distracted according to their interaction log, in the same manner as in \secref{sec:study1-implementation}.
In the \textit{proposed} condition, the ID of the edited slide was sent to a remote server where its continuation was generated by transforming the slide in progress into a text input to the model.
Here, we used the same GPT-3 model as the one used in \secref{sec:study1-implementation} and we never tuned the model for slide-editing.
In addition, we provided part of the model output to a diffusion model~\cite{Rombach_2022_CVPR} to generate an image that complements the generated slide.
The slide was then appended to the original slide via the Google Slides API\footnote{\url{https://developers.google.com/slides/api}}, and the participants were prompted with a notification displaying the title of the generated slide.
In the \textit{control} condition, we used the same six encouraging messages as in \secref{sec:study1-implementation} and showed a notification displaying one of them at the time of distraction.

\subsection{Procedure}
\label{sec:study3-procedure}

We followed the procedure of Study 1, involving twelve participants (eight males and four females; 20--59 years old\footnotemark).
\footnotetext{We note that one participant preferred not to disclose their age.}
Similarly, each participant completed four questionnaires: three at the end of each task and one at the end of all tasks.
From their behavior during slide-editing and their responses to the questionnaires, we prepared the same measures as in \secref{sec:study1-measure}, except for \textit{subjective evaluation} and \textit{progress after resumption}.
For the subjective evaluation, we asked the raters to evaluate each slide with respect to consistency, visual appearance, and overall quality because readability is not applicable to this task.
The progress after resumption is omitted because, at the interface of Google Slide, we could not obtain a precise edit history that allows us to quantify the progress either by the extension or by the API.
Therefore, considering that H2 is supported in \secref{sec:study1-results}, we examined H1, H3, and H4 using the other measures.

\subsection{Results}
\label{sec:study3-results}

\begin{table}[t]
    \centering
    \caption{Comparison of the interest retrieval time between the \textit{proposed} and \textit{control} conditions, which were significantly different ($p = 0.002$).}
    \label{tbl:study3-interest}
    \begin{tabular}{cr@{\hspace{0.5em}}c@{\hspace{0.3em}}r}
        \toprule
                 & mean  &       & \multicolumn{1}{@{\hspace{-0.5em}}c@{\hspace{0.6em}}}{SD} \\
        \midrule
        control  & 49.5  & $\pm$ & 81.8 (s) \\
        proposed & 19.2  & $\pm$ & 44.0 (s) \\
        \bottomrule
    \end{tabular}
\end{table}

\begin{table}[t]
    \centering
    \caption{Comparison of the ignorance rate between the \textit{proposed} and \textit{control} conditions. While the rate was significantly lower for the {proposed} condition ($p < 0.001$), the difference was not observed when we focused on only the first intervention presented to each participant ($p = 0.712)$.}
    \label{tbl:study3-ignorance}
    \begin{tabular}{crrrr}
        \toprule
                 & \multicolumn{2}{c}{(A) \textbf{All}} & \multicolumn{2}{c}{(B) \textbf{First time}} \\
        \cmidrule(lr){2-3} \cmidrule(lr){4-5}
                 & ignored & worked                     & ignored & worked                            \\
        \midrule
        control  & 85      & 104                        & 7       & 11                                \\
        proposed & 19      & 70                         & 4       & 11                                \\
        \bottomrule
    \end{tabular}
\end{table}

\subsubsection{Interest retrieval time}

Regarding H1, we obtained similar results as in \secref{sec:study1-results}; the interest retrieval time was significantly shorter ($t(165.05) = 3.16, p = 0.002$) for the \textit{proposed} condition than for the \textit{control} condition, as presented in \tabref{tbl:study3-interest}.

\subsubsection{Ignorance rate}

Moreover, as shown in \tabref{tbl:study3-ignorance}, the ignorance rate was also significantly lower ($p < 0.001$). 
We also confirmed that, at the first intervention, the \textit{control} condition was considerably effective compared to the \textit{proposed} condition; however, repeated exposure to encouraging messages degraded their effectiveness.

\subsubsection{Total editing time}

For H3, a significant difference was observed in the time the participants spent between the three conditions, according to the Friedman test after Levene's test ($\chi^2(2) = 6.50, p = 0.039$), as shown in \tabref{tbl:study3-time}.
We then conducted the post-hoc Nemenyi test~\cite{nemenyi1963distribution} and found that the time spent in the \textit{proposed} condition was significantly shorter than that in the \textit{none} condition ($p = 0.038$).

\subsubsection{Subjective quality}

On the other hand, we did not observe a significant difference in the raters' evaluation of the quality of the slides the participants edited (\tabref{tbl:study3-quality}), according to the Friedman test ($p = 0.311$, $0.960$, and $0.475$ for consistency, visual appearance, and overall quality, respectively).
Thus, it was implied that the \textit{proposed} condition reduced the total time the participants spent while maintaining the quality of the edited slides.
This suggests that \proposed{} can contribute to improving worker productivity, supporting H3.

\begin{table}[t]
    \centering
    \caption{Comparison of the total editing time between the three conditions. The time spent by the participants in the \textit{proposed} condition was significantly shorter than that in the \textit{none} condition ($p = 0.038$).}
    \label{tbl:study3-time}
    \begin{tabular}{cr@{\hspace{0.5em}}c@{\hspace{0.3em}}r}
        \toprule
                 & mean  &       & \multicolumn{1}{@{\hspace{-0.5em}}c@{\hspace{0.6em}}}{SD} \\
        \midrule
        control  & 1776.1 & $\pm$ &  960.4 (s) \\
        proposed & 1356.5 & $\pm$ & 1048.7 (s) \\
        none     & 2794.5 & $\pm$ & 2866.5 (s) \\
        \bottomrule
    \end{tabular}
\end{table}

\begin{table}[t]
    \centering
    \caption{Comparison of the subjective quality of the slides between the three conditions, which were not significantly different regarding consistency, visual appearance, and overall quality ($p = 0.311$, $0.960$, and $0.475$, respectively).}
    \label{tbl:study3-quality}
    \begin{tabular}{cccc}
        \toprule
                 & consistency     & visual appearance & overall quality    \\
        \midrule
        control  & 5.28 $\pm$ 1.09 & 5.00 $\pm$ 1.15   & 4.92 $\pm$ 1.16  \\
        proposed & 5.22 $\pm$ 1.02 & 5.03 $\pm$ 1.18   & 5.08 $\pm$ 1.11  \\
        none     & 5.42 $\pm$ 0.87 & 4.89 $\pm$ 1.28   & 5.00 $\pm$ 1.20  \\
        \bottomrule
    \end{tabular}
\end{table}

\begin{table}[t]
    \centering
    \caption{Comparison of the participants' evaluations of the system usability between the \textit{proposed} and \textit{control} conditions, which were not significantly different ($p = 0.101$).}
    \label{tbl:study3-sus}
    \begin{tabular}{c@{\hspace{0.1em}}r@{\hspace{0.25em}}c@{\hspace{0.2em}}c@{\hspace{0.7em}}}
        \toprule
                 & \multicolumn{1}{r@{\hspace{-0.2em}}}{mean} &       & SD   \\
        \midrule
        control  & 50.6  & $\pm$ & 19.4 \\
        proposed & 68.8  & $\pm$ & 20.1 \\
        \bottomrule
    \end{tabular}
\end{table}

\begin{figure*}[t]
    \centering
    \includegraphics[width=0.85\linewidth]{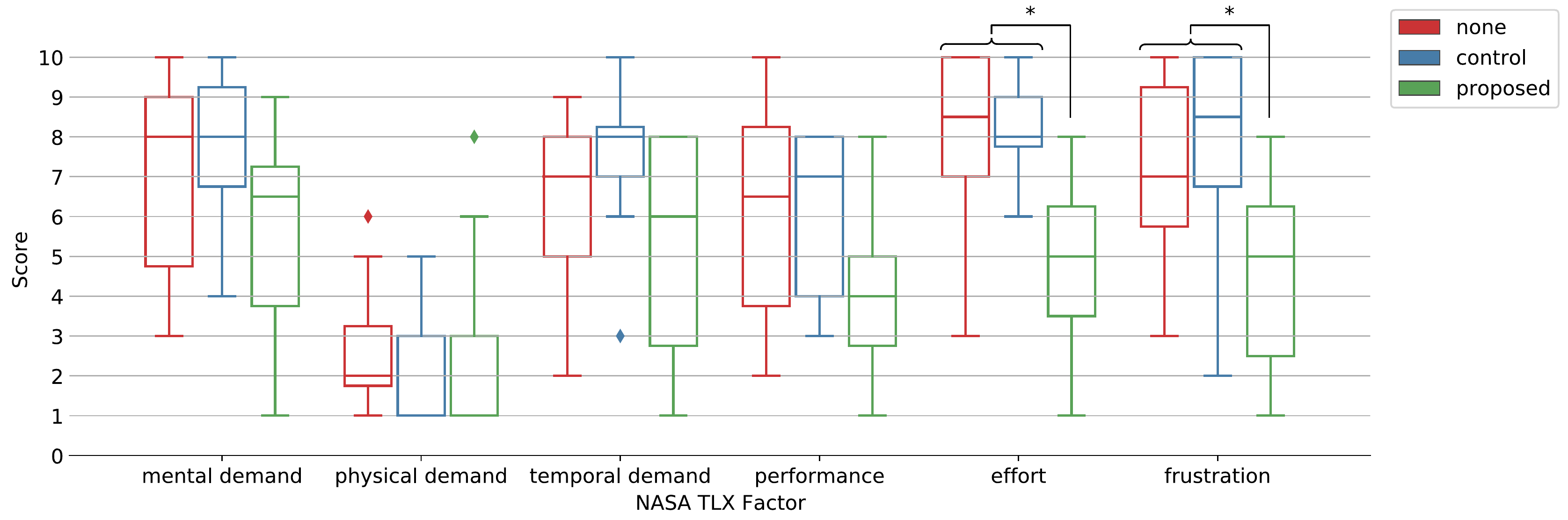}
    \caption{Participants' evaluations of their cognitive load in Study 3. The effort they made and the frustration they felt in the \textit{proposed} condition was significantly lower than those of the \textit{control} ($p = 0.013$ and $0.033$, respectively) and \textit{none} ($p = 0.035$ and $0.033$, respectively) conditions.}
    \Description{This chart shows the result of the cognitive load measured by the NASA-TLX test. There are six factors from left to right: mental demand, physical demand, temporal demand, performance, effort, and frustration. Each with three box plots corresponds to the none, control, and proposed condition. The y-axis shows the score spanning from 0 to 10. For all factors, the proposed condition exhibits a trend of lower scores compared with the other conditions, implying a lower cognitive load. In addition, for the effort and frustration factors, a star illustrating the existence of significant differences in the scores between the proposed condition and the others is shown.}
    \label{fig:study3-tlx}
\end{figure*}

\subsubsection{Cognitive load}

Based on the participants' responses to the questionnaire, we then examined their evaluations of the cognitive load.
As illustrated in \figref{fig:study3-tlx}, we found significant differences in cognitive load in terms of temporal demand ($F(2, 22) = 3.99, p = 0.033$), performance ($F(2, 22) = 3.79, p = 0.038$), effort ($F(2, 22) = 8.63, p = 0.001$), and frustration ($F(2, 22) = 7.80, p = 0.003$) according to ANOVA.
The post-hoc test with Holm correction revealed that the scores of the effort and frustration factors in the \textit{proposed} condition were significantly lower than those in the \textit{control} ($t(11) = 3.60, p = 0.013$ and $t(11) = 3.04, p = 0.033$, respectively) and \textit{none} ($t(11) = 2.79, p = 0.035$ and $t(11) = 2.9, p = 0.033$, respectively) conditions.

These results implied the advantage of our design.
That is, the design of guiding the participants' interest using the generated continuation of the interrupted work, which is context-aware and variational, would ease their frustration.
This would reduce the effort required to resume the task as well.
Simultaneously, as mentioned in \secref{sec:study1-comments} and also later in \secref{sec:study3-comments}, the generated continuation sometimes supported their progress by providing some ideas or reference information.
This corroborated our observation that \proposed{} contributes to participants' perceived utility.

\subsubsection{System usability}

The participants' responses to items that measured system usability further supported this observation.
As presented in \tabref{tbl:study3-sus}, the participants provided a higher SUS score for the \textit{proposed} condition than for the \textit{control} condition on average, although their difference was insignificant ($t(11) = 1.79, p = 0.101$).
Furthermore, the three conditions exhibited a significant difference in the participants' rankings regarding usability, according to the Friedman test ($\chi^2(2) = 7.09, p = 0.029$).
We found that the \textit{proposed} condition was likely to be ranked significantly higher than the \textit{control} condition ($p = 0.038$), as eight participants regarded the \textit{proposed} condition as the best.
This result supported H4, and together with the above results, we conclude that the effectiveness and extensibility of \proposed{} were confirmed.

\subsection{User Comments}
\label{sec:study3-comments}

The comments of the participants highlighted the advantage of \proposed{} in drawing their interest and lowering the hurdle for resumption in the slide-editing task as well.
\begin{quote}
    I was surprised at the high quality of the automatically generated slides. I used the presented images as is and the provided text with a little editing.
\end{quote}
\begin{quote}
    When I was notified with just a message, I was a little uncomfortable because I felt like being monitored. Regarding the notification from the AI, I was able to regard it as if the AI tried to give me an idea, so it was more satisfactory.
\end{quote}
These comments are consistent with their evaluation of cognitive load.

However, some participants pointed out the limitation of the generated continuation, which might reflect the fact that the model used was not specifically tuned for slide-editing.
\begin{quote}
    As the quality of the AI's output was not always perfect, the process of deleting the automatically added slides or extracting useful parts from the slides and merging them into my own slides was somewhat laborious and demanding.
\end{quote}
Still, the participant commented that ``it was very helpful that the AI provided content ideas, which allowed me to focus consistently on the task without wondering.''
In other words, it was implied that large generative models that are not tuned for each individual task and might be imperfect can still be useful in helping workers avoid procrastination.

\section{Findings and Discussion}

Thus far, we have demonstrated the effectiveness and domain extensibility of \proposed{} by verifying our hypotheses through two different tasks (writing and slide-editing).
Specifically, the results demonstrated \proposed{} can help workers avoid task procrastination with less cognitive load through interventions made by generative models without fine-tuning.
At the same time, the series of studies informed us of further linkages between our study and prior work on human-AI collaborations.
In particular, we found that the results highlighting the importance of reframing the role of generative models to induce workers' behavior change, rather than directly contributing to their tasks, corroborate the previous discussion.
In this section, we discuss the findings of our study to clarify the linkages.

\subsection{Exploiting Large Generative Models without Fine-tuning}

Recent advances in large generative models are remarkable, such that they allow workers to delegate more work to AIs, raising expectations for achieving ``the AI leads and the human assists'' or even ``full AI automation''~\cite{DBLP:conf/nips/LubarsT19}. 
Some of our results reconfirmed this capability; for instance, some participants appreciated the high-quality content generated by \proposed{} and actually incorporated it.
Concurrently, the results highlighted the imperfections of the models and how they affect workers' experience, as the participants mentioned their insufficient quality in various writing domains in \secref{sec:study2-results-accuracy}.
Even if the generation techniques continue to advance, it is not guaranteed that they work perfectly in diverse task domains. 
Thus, the wall of accuracy will remain in individual workers' tasks, which can be an inevitable issue when we aim for a higher level of task delegation; reported failure cases often include degraded trust and ignorance of the systems~\cite{DBLP:conf/iui/YangZZLL22, DBLP:conf/iui/KimMS21, DBLP:conf/chi/YanCKGAH22}.
Tuning generative models to achieve higher accuracy is a theoretically valid solution; however, it is impractical because of its tremendous cost~\cite{DBLP:conf/aaai/StrubellGM20, DBLP:journals/corr/abs-2007-03051}, especially considering that there are many tasks and domains that each worker needs support for.
Moreover, small datasets are often inadequate for tuning such large models~\cite{DBLP:conf/naacl/DevlinCLT19, DBLP:journals/corr/abs-1811-01088}, which makes the solution intractable.

\proposed{} demonstrated an alternative way of contributing to the digital well-being of workers.
In other words, our motivation for using generative models is different in the sense that it does not aim to increase task delegation but to influence workers' behavior.
Our results exhibited that the ability of large generative models to produce context-aware content can be utilized for such interventions.
More importantly, the results demonstrated its applicability in multiple domains without fine-tuning, suggesting a novel way of leveraging publicly available large models, in a bricolage manner, to support individual workers.

\subsection{Passive Intervention vs. Active Use}
\label{sec:disc-passive}

An interesting conflict implied from the participants' responses in \secref{sec:study2-results-improvement} is that some participants requested a button so that they could use AIs actively, while others mentioned concerns about the idea.
This discrepancy can be explained from the perspective of the relationship between expectations and trust toward AIs.
As Yin \etal~\cite{DBLP:conf/chi/YinVW19} suggested, workers' trust in the system would be affected by how much they expect in the system and how accurate the observed output is.
In this sense, placing a button for actively using AIs can lead to workers' raised expectations, as they would rely on the feature to boost their work, and thus, their trust could be degraded if the output quality is insufficient.
Therefore, in our design of \proposed{}, we did not include such buttons and limited the role of AIs to passive intervention for workers only when they were detected to be distracted.
This idea aligns with the concept of \textit{unremarkable AI} proposed by Yang \etal~\cite{DBLP:conf/chi/YangSZ19}.
They designed a clinical decision support tool in which the existence of AI can only be noticed when it adds value to the decision while maintaining the existing decision-making routine.
Similarly, in our case, we did not introduce generative models as \textit{remarkable AI} within workers' task routines.
We rather limited their presence outside of the original task (\ie when they are away from the task), which was supported by the participants.
Simultaneously, their comments asking for buttons motivated us to conduct further research to examine the possibility of integrating the design of remarkable AI, while maintaining the current advantage of unremarkable AI.
In addition, it is also possible to incorporate a design where generated continuations are continuously shown to workers and they can select whether to use them. 
We anticipate that such mixed-initiative interaction can lead to even better human-AI collaboration, as Lehmann and Buschek~\cite{DBLP:journals/icom/LehmannB21} discussed using the concept of autocomplete.

\section{Future Work and Limitations}

Even though we verified the effectiveness of \proposed{} in two different tasks (writing and slide-editing), further research is demanded to test its domain extensibility.
As we mentioned in \secref{sec:rw-generative}, generative models have been proposed for other tasks, such as drawing.
The advantage of \proposed{}, that is, its ability to leverage such models without tuning to individual tasks, motivated us to conduct studies in diverse situations involving workers with different proficiency levels and backgrounds.

We also want to explore how users of \proposed{} can incorporate the content of the generated continuations into their final outputs.
This includes several aspects, from directly using sentences or images provided by the system to taking inspiration about possible ideas from them.
Thus, the exploration would require an in-depth observation of the usage of \proposed{} alongside the current series of experiments, which will provide insightful discussions in combination with the findings of Lee \etal~\cite{DBLP:conf/chi/0002LY22}.

Moreover, while we used encouraging messages adopted from a public site blocker for the control condition, further comparison with strategies of other prior work will provide more insights into the relationship between interventions and workers' behavior.
For example, HabitLab~\cite{DBLP:journals/pacmhci/KovacsWB18} provides a rotation of diverse interventions to help users get back on task while maintaining their interest.
It is, thus, a promising approach to use \proposed{}'s intervention as one kind in the rotation to keep the novelty of the intervention.

One limitation is that our quantitative results are limited to laboratory studies in both tasks, similar to prior work~\cite{DBLP:conf/cscw/LiuJPP14}.
This is due to the difficulty in recording the measures we used in a study conducted in the wild and making a fair comparison.
For example, each worker would have tasks with different difficulties and priorities (\eg deadline, meeting, \textit{etc}.).
Therefore, in our in-the-wild study (\secref{sec:study2}), we focused on qualitatively confirming the usefulness of \proposed{}.
It is desirable to conduct a longitudinal study measuring the digital well-being of workers as an outcome, as Howe and Menges did~\cite{doi:10.1080/07370024.2021.1987238}, and quantifying the effectiveness of \proposed{} for future work.

The mechanism of detecting workers' distractions can be further improved based on participants' feedback.
The participants noticed that interventions sometimes occurred, even when they focused on the task.
This is because our na\"{i}ve method detected the status of workers based on the interaction log.
Such methods are unable to consider moments when workers contemplate deeply (\eg thinking about a plot when writing a novel) without making any interactions on the web page.
If we could implement software installed on their PCs, instead of a prototype web page, it would be possible to monitor more fine-grained activities, not just typing or not on a specific web page.
Such information will help detect moments when workers are actually distracted with higher precision~\cite{DBLP:conf/icmi/ThomasJ17,DBLP:journals/umuai/YakuraNG22}, which would also contribute to quantifying the effectiveness of \proposed{} by contextualizing the moment in the longitudinal in-the-wild study mentioned above.
Still, we want to emphasize that \proposed{} did not degrade workers' experiences compared to conventional encouraging messages through false-positive interventions.
This is because \proposed{} can offer hints to foster their contemplation rather than alert them.
Such false-positive resistance is an important aspect for machine-learning-based intervention to be successful~\cite{DBLP:conf/chi/ArakawaY21} and exactly what \proposed{} aimed at, as mentioned in \secref{sec:proposed-description-detection}.

We also note that the use of large generative models poses a risk of bringing about some bias and ethical considerations~\cite{DBLP:journals/csur/MehrabiMSLG21}.
For example, stereotypes and negative propagation of particular social groups have been observed in their outputs~\cite{DBLP:conf/emnlp/ShengCNP19, DBLP:conf/fat/BenderGMS21, DBLP:journals/corr/abs-2202-04053}.
In \proposed{}, we cannot deny the possibility that such a bias is reflected in the final production of workers via the generated continuation.
Toward deployment of \proposed{}, it is desirable to integrate measures to mitigate such bias~\cite{DBLP:journals/csur/MehrabiMSLG21} as well as provide a transparent explanation of the used models~\cite{DBLP:conf/fat/MitchellWZBVHSR19} and the clarification of the risk.

\section{Conclusion}

We presented a novel approach, \proposed{}, for supporting office workers in their task engagement through an intervention that leverages large generative models.
Specifically, \proposed{} generates a continuation of interrupted work to prompt workers to resume the task by effectively attracting their interest and inducing behavior change for task resumption.
The approach is domain-extensible, which means that we can develop systems on top of \proposed{} using publicly available models without having to tune them to individual tasks.
We conducted a series of studies to evaluate the effectiveness and usability of \proposed{}.
The first study confirmed the effectiveness of \proposed{} in a writing task compared to a conventional encouraging approach.
The second study further supported the favorable usability of \proposed{} throughout an in-the-wild study over a longer period by involving participants with different purposes for writing.
The third study confirmed the effectiveness of \proposed{} in a slide-editing task, demonstrating its domain extensibility.
Based on the results of the studies, we discussed the implications of our design, that is, leveraging large generative models to influence workers' behavior instead of delegating human tasks to AIs.
Our findings and discussion pave the way for developing novel human-AI collaborations to enhance workers' digital well-being that can be harnessed in future human-AI symbiosis.

\begin{acks}
This work was supported in part by JST ACT-X Grant Number JPMJAX200R and JSPS KAKENHI Grant Numbers JP21J20353.
\end{acks}

\bibliographystyle{ACM-Reference-Format}
\bibliography{paper}

\end{document}